%% file: Manuscript.tex
\documentclass[preprint,12pt]{elsarticle}

\input{./utils/imports}
\input{./utils/macros}
\input{./utils/mute-warnings}

\journal{Computer Physics Communications}

\begin{document}

\begin{frontmatter}



\title{Mumax3-cQED: an extension of Mumax3 to simulate magnon-photon interactions in cavity QED }


\author[inma-addr]{Sergio Mart\'inez-Losa del Rinc\'on\corref{author1}\fnref{equal}}
\author[inma-addr]{Juan Rom\'an-Roche \fnref{equal}}
\author[inma-addr]{\\Andr\'es Mart\'in-Megino}
\author[inma-addr]{David Zueco\fnref{author4}}
\author[inma-addr]{\\Mar\'ia Jos\'e Mart\'inez-P\'erez\fnref{author5}}

\fntext[equal]{These authors contributed equally to this work.}
\fntext[author4]{\linkemail{dzueco@unizar.es}}
\fntext[author5]{\linkemail{pemar@unizar.es}}

\address[inma-addr]{Instituto de Nanociencia y Materiales de Aragón (INMA), CSIC-Universidad de Zaragoza, Zaragoza, ES-50009 Spain}

\begin{abstract}
We present an extension of the well-known micromagnetic package Mumax3 to simulate magnon-polaritons in realistic magnetic materials and nanostructures. Mumax3-cQED leverages the full GPU-accelerated capabilities of Mumax3 to model standard spin-spin interactions and the coupling of magnetic moments to external space- and time-dependent magnetic fields, with the additional unique feature of including the coupling to a cavity. We validate the code against results obtained from the Dicke model in both the  paramagnetic and the superradiant  phases. We show that hybrid magnon-light states can be calculated, as well as the non-equilibrium dynamics and their approach to equilibrium. In addition, we demonstrate the potential of Mumax3-cQED to reproduce experimental results and design magnon-cavity experiments, including three-dimensional and coplanar waveguide resonators. The code is fully available and will be useful for designing experiments involving microscopic saturated ferromagnets as well as systems featuring spin textures such as domain walls, vortices, or skyrmions.
\end{abstract}


\end{frontmatter}



\noindent{\bf PROGRAM SUMMARY/NEW VERSION PROGRAM SUMMARY}

\begin{small}
\noindent
{\em Program Title:} MUMAX3-CQED \\
{\em CPC Library link to program files:} (to be added by Technical Editor) \\
{\em Developer's repository link:} (if available) \\
{\em Code Ocean capsule:} (to be added by Technical Editor)\\
{\em Licensing provisions:} GPLv3 \\
{\em Programming language:} Go, CUDA, C, Makefile, Shell, PowerShell \\
%
\end{small}

%
\input{./sections/00-introduction} 

\input{./sections/01-llg-cavity}

\input{./sections/02-mumax-modification}

\input{./sections/03-benchmarking-mumax-cqed}

\input{./sections/04-nanowire-cavity}

%

\appendix

\input{./sections/appendix_i}



\bibliographystyle{elsarticle-num}
\bibliography{Manuscript}







\end{document}


%% file: utils/imports.tex
\usepackage[utf8]{inputenc}
\usepackage[T1]{fontenc}
\usepackage{listings}
\usepackage{etoolbox}
\usepackage{xfrac}
\usepackage{atbegshi}
\usepackage{atveryend}
\usepackage[labelfont=bf]{caption}
\usepackage{subcaption}
\usepackage{adjustbox}
\usepackage{empheq}
\usepackage{amsmath}
\usepackage{ulem}
\usepackage{graphicx}
\usepackage{amssymb}

\biboptions{super,square,comma}

\usepackage{tikz}
\usetikzlibrary{shapes.geometric, arrows, calc, 3d}
\usepackage{relsize}
\usepackage[most]{tcolorbox}
\usepackage[frozencache=true,cachedir=minted-cache]{minted}
\usepackage{hyphenat}
\usepackage{ifthen}
\usepackage{./packages/catoptions}
\usepackage{algorithm}
\usepackage{algpseudocode}
\usepackage{listings}
\usepackage{xcolor}
\usepackage{hyperref}

\definecolor{color-hyperref}{HTML}{436491}

\hypersetup{
	colorlinks = true,
	linkcolor=color-hyperref,
	citecolor=color-hyperref,
	filecolor=color-hyperref,
	urlcolor=color-hyperref,
	runcolor=color-hyperref,
	menucolor=color-hyperref,
	linkbordercolor=color-hyperref,
	citebordercolor=color-hyperref,
	filebordercolor=color-hyperref,
	urlbordercolor=color-hyperref,
	runbordercolor=color-hyperref,
	menubordercolor=color-hyperref,
	pdfpagemode=UseOutlines,
	hypertexnames = true,
	pdfencoding = auto, 
	psdextra, 
	bookmarksdepth = 4
	bookmarks=true,
	bookmarksopen=true,    
	bookmarksopenlevel=0,
	bookmarksnumbered=true,
	plainpages=false
}

\graphicspath{{./images/}}

\DeclareGraphicsExtensions{.eps,.png,.pdf,.jpg}


%% file: utils/macros.tex
\newcounter{bla}

\newcommand{\italic}[1]{\textit{#1}}

\newrobustcmd*{\linkemail}[1]{\textit{Email address:} \mail{#1}}
\newrobustcmd*{\mail}[1]{\href{mailto:#1}{\ttfamily\bfseries#1}}

\makeatletter

\newcommand\@ifnextchars[3]{%
	\def\@tempa{#1}\def\@tempb{#2}%
	\def\@tempc{#3}\def\@tempd{}%
	\@ifnextchars@a%
}%

\def\@ifnextchars@a#1{%
	\toks@\cptthreexp{\expandafter\@car\@tempa\@nil}%
	\expandafter\ifstrcmpTF\expandafter{\the\toks@}{#1}{%
		\edef\@tempd{\expandcsonce\@tempd\the\toks@}%
		\toks@\cptthreexp{\expandafter\@cdr\@tempa\@nil}%
		\edef\@tempa{\the\toks@}%
		\ifcsemptyTF\@tempa{\expandafter\@tempb\@tempd}%
		\@ifnextchars@a
	}{%
		\edef\@tempd{\expandcsonce\@tempd#1}%
		\expandafter\@tempc\@tempd
	}%
}

\makeatother

\newcommand{\super}[1]{\textsuperscript{#1}}

\makeatletter

\DeclareRobustCommand{\checkWord}[1]{\@ifnextchars{,}{\expandafter#1}{\@ifnextchars{)}{\expandafter#1}{\@ifnextchars{.}{\expandafter#1}{\expandafter#1\hspace{0.3em}}}}}

\def\mumaxText{mumax\super{3}}
\def\MumaxText{Mumax\super{3}}
\def\MUMAXText{MUMAX\super{3}}

\def\mumaxmantisa{\texorpdfstring{\textsuperscript{3}\space}{\textthreesuperior\space}}%

\def\mumaxTitle{mumax\expandafter\mumaxmantisa}
\def\MumaxTitle{Mumax\expandafter\mumaxmantisa}
\def\MUMAXTitle{MUMAX\expandafter\mumaxmantisa}

\def\Name#1#{%
	\ifx\\#1\\%
		\expandafter\@firstoftwo%
	\else%
		\expandafter\@secondoftwo%
	\fi%
}%

\newcommand*{\Namemumax}{%
	mumax\expandafter\mumaxmantisa%
}%

\newcommand*{\NameMumax}{%
	Mumax\expandafter\mumaxmantisa%
}%

\newcommand*{\NameMUMAX}{%
	MUMAX\expandafter\mumaxmantisa%
}%

\newcommand{\mumax}[1][1]{%
	\ifthenelse{\equal{#1}{1}}{%
		\checkWord{\mumaxText}%
	}{%
		\ifthenelse{\equal{#1}{2}}{%
			\checkWord{\MumaxText}%
		}{%
			\ifthenelse{\equal{#1}{3}}{%
				\mumaxText%
			}{%
				\checkWord{\MUMAXText}%
			}%
		}%
	}%
}%

\def\ps@pprintTitle{%
	\let\@oddhead\@empty
	\let\@evenhead\@empty
	\def\@oddfoot
	{\hbox to \textwidth%
		{\ifnopreprintline
			\relax
		 \else
			\@myfooterfont%
			\ifx\@elsarticlemyfooteralign\@elsarticlemyfooteraligncenter%
				\hfil\@elsarticlemyfooter\hfil%
			\else%
				\ifx\@elsarticlemyfooteralign\@elsarticlemyfooteralignleft%
					\@elsarticlemyfooter\hfill{}%
				\else%
					\ifx\@elsarticlemyfooteralign\@elsarticlemyfooteralignright%
						{}\hfill\@elsarticlemyfooter%
					\else%
						Preprint submitted to %
						\ifx\@journal\@empty%
							Elsevier%
						\else%
							\@journal\fi\hfill\@date%
						\fi%
					\fi%
			 	\fi%
			 \fi%
		}%
		\raise -6.28ex%
		\hbox to -\textwidth{%
			\hss\thepage\hss%
		}%
	}%
	\let\@evenfoot\@oddfoot
}

\tikzstyle{startstop} = [rectangle, rounded corners, 
							minimum width=3cm, 
							minimum height=1cm,
							text centered, 
							draw=black, 
							fill=red!30]

\tikzstyle{io} = [trapezium, 
					trapezium stretches=true,
					trapezium left angle=70, 
					trapezium right angle=110, 
					minimum width=3cm, 
					minimum height=1cm, text centered, 
					draw=black, fill=blue!30]

\tikzstyle{process} = [rectangle, 
						minimum width=3cm, 
						minimum height=1cm, 
						text centered, 
						text width=3cm, 
						draw=black, 
						fill=orange!30]

\tikzstyle{decision} = [diamond, 
						minimum width=3cm, 
						minimum height=1cm, 
						text centered, 
						draw=black, 
						fill=green!30]
						
\tikzstyle{arrow} = [thick,->,>=stealth]

\definecolor{LightGrayCode}{gray}{0.96}

\newtcbox{\mathbox}[1][]{%
	nobeforeafter, math upper, tcbox raise base,%
	enhanced, colframe=black!30!black,%
	colback=black!30, boxrule=1pt,%
	#1}%

\makeatletter

\def\footnotecolour{\color{red}}

\renewcommand\@makefnmark{\hbox{\@textsuperscript{\normalfont\footnotecolour\@thefnmark}}}
\renewcommand\@makefntext[1]{%
	\parindent 1em\noindent
	\hb@xt@1.8em{%
		\hss\@textsuperscript{\normalfont\@thefnmark}}#1}
\makeatother

\def\spaceBelowCaptionSubfigures{6pt}
\def\spaceBelowCaptionFigures{-6pt}
\def\spaceAboveCaptionSpace{6pt}

\newcommand{\imagefigurecaption}[4][0.35]{%
	\begin{figure}[H]
		\centering
		\includegraphics[width=#1\textwidth,height=\textheight,keepaspectratio]{#2}
		\captionsetup{aboveskip=\spaceAboveCaptionSpace, belowskip=\spaceBelowCaptionFigures, justification=raggedright, singlelinecheck=true}
		\caption{#3}
		\ifthenelse{\equal{#4}{}}{}{\label{#4}}
	\end{figure}	
}

\DeclareRobustCommand\nocitecaptionlink[1]{{\def\cite##1{}#1}}%
\newcommand\nocitecaption[1]{\caption[\nocitecaptionlink{#1}]{#1}}%

\newcommand{\imageinsets}[9]{%
	\begin{figure}[H]
		\centering
		\begin{tikzpicture}[every node/.style={anchor=south west,inner sep=0pt}, x=1mm, y=1mm]
			\node (image1) at (0,0) {\includegraphics[width=#3\textwidth,height=\textheight,keepaspectratio]{#1}};
			\node[path picture={
				\fill[#6, opacity=#7](path picture bounding box.north west)rectangle
				($(path picture bounding box.north east)!1.0!(path picture bounding box.south east)$);
			}] (image2) at #5 {\includegraphics[width=#4\textwidth,height=\textheight,keepaspectratio]{#2}};
		\end{tikzpicture}%
		\captionsetup{aboveskip=\spaceAboveCaptionSpace, belowskip=\spaceBelowCaptionFigures, justification=raggedright, singlelinecheck=true}
		\ifthenelse{\equal{#8}{}}{}{\caption{#8}}
		\ifthenelse{\equal{#9}{}}{}{\label{#9}}
	\end{figure}%
}%

\newcommand{\centercaption}[1]{	
	\captionsetup{aboveskip=\spaceAboveCaptionSpace, belowskip=\spaceBelowCaptionFigures, justification=justified, singlelinecheck=true}
	\protect\caption{#1}
}

\newcommand{\nocitecentercaption}[1]{	
	\captionsetup{aboveskip=\spaceAboveCaptionSpace, belowskip=\spaceBelowCaptionFigures, justification=raggedright, singlelinecheck=true}
	\protect\nocitecaption{#1}
}

\makeatletter
\renewcommand{\p@subfigure}{\thefigure.}
\makeatother

\newcommand{\asideimagescontinue}[3]{%
	\begin{figure}[H]
		\ifthenelse{\equal{\tempasidee}{}}{}{
			\ifthenelse{\equal{\tempasideg}{}}{}{
				\captionsetup[subfigure]{belowskip=\spaceBelowCaptionSubfigures, labelformat=empty, justification=raggedright, singlelinecheck=true}
			}
		}
		\begin{subfigure}{.48\textwidth}
			\centering
			\includegraphics[width=\tempasideb]{\tempasided}
			\ifthenelse{\equal{\tempasidee}{}}{}{\caption{\tempasidee}}
			\ifthenelse{\equal{#2}{}}{}{\label{#2}}
		\end{subfigure}%
  		\hspace{0.8em}
		\begin{subfigure}{.48\textwidth}
			\centering
			\includegraphics[width=\tempasidec]{\tempasidef}
			\ifthenelse{\equal{\tempasideg}{}}{}{\caption{\tempasideg}}
			\ifthenelse{\equal{#3}{}}{}{\label{#3}}
		\end{subfigure}%
		\ignorespaces\lowercase{\def\tmpVal{\tempasidea}}
		\ifthenelse{\equal{\tmpVal}{false}}{
			\centercaption{\tempasideh}
		}{	
			\nocitecentercaption{\tempasideh}
		}
		\label{#1}
	\end{figure}
}%

\newcommand{\tikzimagefigurelabel}[3]{%
	\begin{figure}[H]%
		\centering%
		#2%
		\captionsetup{belowskip=\spaceBelowCaptionFigures, justification=raggedright, singlelinecheck=true}%
		\caption{#3}%
		\ifthenelse{\equal{#1}{}}{}{\label{#1}}%
	\end{figure}%
}

\definecolor{folderbg}{RGB}{124,166,198}
\definecolor{folderborder}{RGB}{110,144,169}

\newlength\Size
\tikzset{%
	folder/.pic={%
		\filldraw [draw=folderborder, top color=folderbg!50, bottom color=folderbg] (-1.05*\Size,0.2\Size+5pt) rectangle ++(.75*\Size,-0.2\Size-5pt);
		\filldraw [draw=folderborder, top color=folderbg!50, bottom color=folderbg] (-1.15*\Size,-\Size) rectangle (1.15*\Size,\Size);
	},
	file/.pic={%
		\filldraw [draw=folderborder, top color=folderbg!5, bottom color=folderbg!10] (-\Size,.4*\Size+5pt) coordinate (a) |- (\Size,-1.2*\Size) coordinate (b) -- ++(0,1.6*\Size) coordinate (c) -- ++(-5pt,5pt) coordinate (d) -- cycle (d) |- (c) ;
	},
}

\makeatletter
\renewcommand{\ALG@beginalgorithmic}{\footnotesize}
\makeatother



\definecolor{codegreen}{rgb}{0,0.6,0}
\definecolor{codegray}{rgb}{0.5,0.5,0.5}
\definecolor{codepurple}{rgb}{0.58,0,0.82}
\definecolor{backcolour}{rgb}{0.95,0.95,0.92}

\lstdefinestyle{script-style}{
    backgroundcolor=\color{backcolour},   
    commentstyle=\color{codegreen},
    keywordstyle=\color{magenta},
    numberstyle=\tiny\color{codegray},
    stringstyle=\color{codepurple},
    basicstyle=\ttfamily\footnotesize,
    breakatwhitespace=false,         
    breaklines=true,                 
    captionpos=b,                    
    keepspaces=true,                 
    numbers=left,                    
    numbersep=5pt,                  
    showspaces=false,                
    showstringspaces=false,
    showtabs=false,                  
    tabsize=2
}

%% file: utils/mute-warnings.tex
\usepackage[immediate]{silence} 

\newcount\hbadness
\newdimen\hfuzz

%% file: sections/00-introduction.tex
Bondarenko2023
Hrabec2020 synthetic skyr
Valenzuela2024   spintronics YA
Bourhill2023 circulating magn YA
Lee2023 nonlinear magnon  YA
Bondarenko2023  vortex coupling ttransductoin mumax YA
Sharma2022 quantum state qubit cavity magnet ya

\section{Introduction}
\label{sec:introduction}

Cavity quantum electrodynamics (cQED) deals with the light-matter interactions in the regime where the discrete nature of light is relevant\cite{Walther2006}. This has important potential applications in the fields of quantum computing, communication, sensing and metrology\cite{Blais2021}. On the other hand, cQED allows testing fundamental predictions of quantum mechanics, e.g., enhanced spontaneous emission\cite{Purcell1946};  shifts of the energy levels\cite{Brune1994} and  Rabi oscillations\cite{Johansson2006}; or superradiance\cite{Hepp1973}.  
In the so-called strong coupling regime, light-matter hybrid states, known as polaritons, can be resolved experimentally\cite{Basov2020}. 

Many different polaritons have been observed  including (artificial) spins and collective matter excitations such as phonons, plasmons or magnons, i.e. quanta of spin waves in ferromagnets.
Magnons and their interactions are a hot topic of research due to their potential in developing quantum devices\cite{Jiang2023,ZareRameshti2022,LachanceQuirion2019,Chumak2015}, including optical-to-microwave transducers\cite{Bondarenko2023,Lambert2019,Hisatomi2016}, dark matter axion detectors\cite{Barbieri1989, Flower2019}, and quantum memories\cite{Li2022,Zhang2015}. Beyond these practical applications, magnons feature unique properties such as their extremely small wavelength (even at microwave frequencies\cite{GonzalezGutierrez2024}), their gyrotropic dynamics (which naturally break time-reversal symmetry\cite{Asadchy2020,Wang2019}), and their potential to create non-Hermitian systems (enabling the observation of exceptional points\cite{Cao2019,Zhang2017} and level attraction\cite{Harder2018,Harder2021}). Most of these phenomena are based on the combination of magnons and cavity photons, making it particularly important to understand, build and control magnon-polaritons.


%
\begin{figure}
    \centering
    \includegraphics[width =0.7\columnwidth]{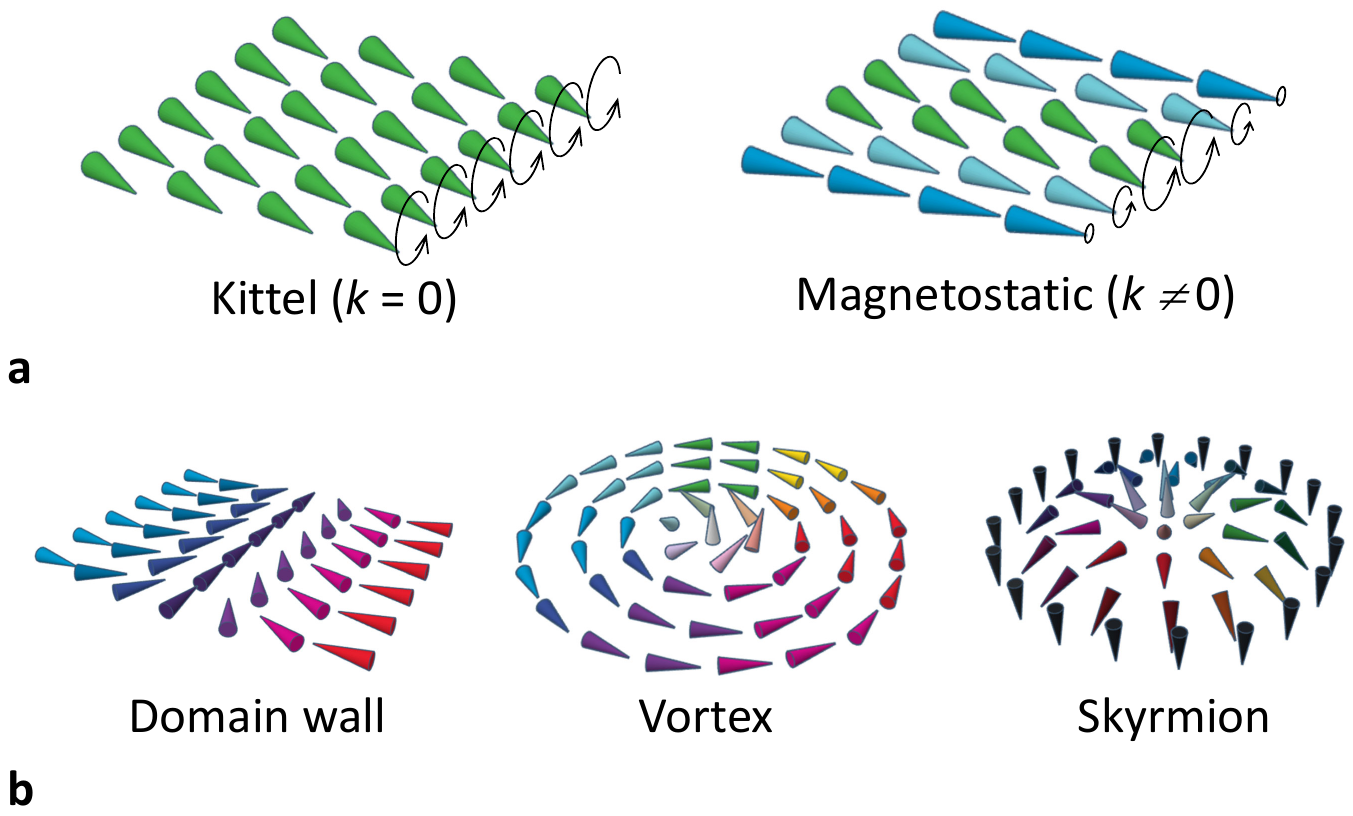}
    \caption{a: Spin-wave modes in saturated ferromagnets. All spins precede at unison in the homogeneous Kittel mode while the amplitude of spin precession varies along the ferromagnet in magnetostatic higher-$k$ modes. The latter can be excited using non-homogeneous microwave magnetic fields.  b: Ground states of different spin textures.}
    \label{fig:Fig1}
\end{figure}
%


There are different theoretical methods\cite{Joseph2024,MartinezLosadelRincon2023,Lee2023,Macedo2021,Bourhill2020,Proskurin2019,Flower2019a,MartinezPerez2018,Goryachev2014,Soykal2010} that provide good estimations of the magnon-photon coupling strength in saturated ferromagnets, where the homogeneous (Kittel, $k=0$) or high-$k$ (magnetostatic) modes can be excited (see Fig. \ref{fig:Fig1}a). However, this is only true under the assumption that the magnetic susceptibility of the ferromagnet can be calculated. In other words, a precise knowledge of the demagnetizing factors is needed. Unfortunately, the latter can be calculated exactly only for ellipsoids of revolution. In contrast, the ground state of real ferromagnets is typically non homogeneous, as it results from the competition between a number of energies including Zeeman, exchange, dipolar and magnetocrystalline anisotropies. Non-homogeneous demagnetizing fields typically yield the excitation of high-$k$ modes. Under certain conditions, these contributions can even yield magnetic textures\cite{Yu2021}, as summarized in Fig. \ref{fig:Fig1}b. For example: domain walls separating regions with different magnetization direction; curls of the magnetization referred as vortices; and magnetic skyrmions, i.e., topologically protected defects stabilized in materials featuring asymmetric exchange interactions. The dynamic evolution of these objects cannot be calculated analytically and micromagnetic computation is typically needed\cite{Leliaert2019}. Numerical codes such as OOMMF or Mumax3 are widely used due to the many applications of magnetic textures in spintronics\cite{Valenzuela2024,Yu2021}. 



%
\begin{figure}
    \centering
    \includegraphics[width =0.7\columnwidth]{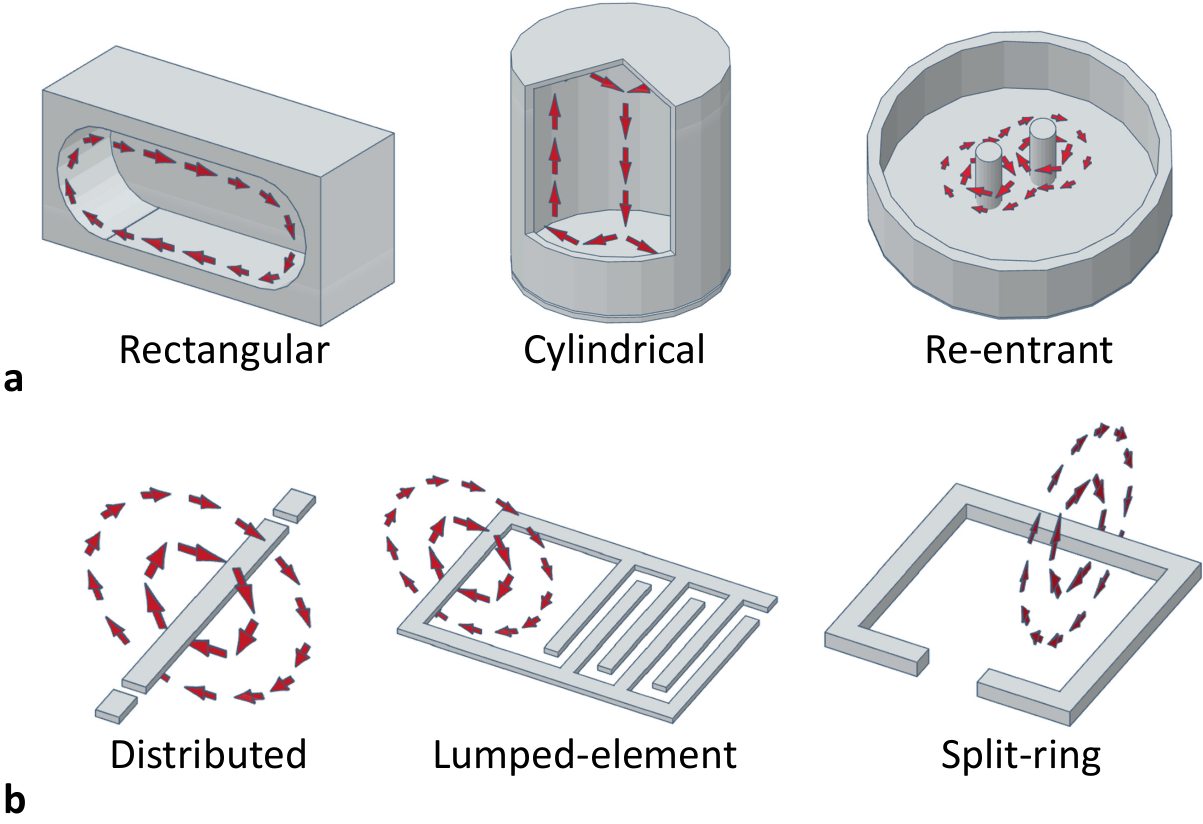}
    \caption{a: Different implementations of metallic three-dimensional electromagnetic cavities. b: Cpw resonator configurations, usually fabricated out of superconductors. In all panels, one magnetic mode is represented by red arrows.}
    \label{fig:Fig2}
\end{figure}

To make things even more complicated, electromagnetic cavities come in a number of ways, starting from the early implementations using Fabry-Perot mirrors\cite{Haroche2020}, up to solid-state cavity implementations (see Fig. \ref{fig:Fig2}a). Experiments on cavity magnonics benefit from the use of (quasi)homogeneous microwave field configurations in three dimensional cavities that only couple to the Kittel mode in magnetic spheres\cite{Tabuchi2014}. By increasing the size of the sample, the ultra-strong coupling regime has been observed, while also coupling  to higher order magnetostatic modes due to in-homogeneities in the microwave field across the volume of the sphere\cite{Bourhill2016,Zhang2014}. Tuneable cylindrical cavities have been used to observe anti-resonant modes that yield coherent (level-repulsion)  or dissipative (level-attraction) magnon-photon couplings\cite{Rao2019}. Using multiple excitation ports, microwave fields with definite polarization can be produced. This allows for  selective coupling to Kittel modes with fixed gyration set by the external field \cite{Joseph2024}, or tuning between the regimes of coherent and dissipative coupling \cite{Boventer2019}. Re-entrant cavities with multiple number of posts are very appealing since they allow to focus the microwave magnetic mode into small regions to achieve ultra-strong couplings\cite{Flower2019a}. Moreover, combining different posts it is possible to obtain highly non-homogeneous microwave configurations that couple selectively to precise magnetostatic modes\cite{Gardin2024,Bourhill2023,Owens2022,Goryachev2014}. Moving from three dimensions to planar cavities is also of extreme relevance to build on-chip quantum devices (see Fig. \ref{fig:Fig2}b). Starting from the first experimental observation of strong magnon-photon coupling \cite{Huebl2013} to recent experiments even reaching the ultra-strong \cite{Ghirri2023} or approaching the deep-strong \cite{Golovchanskiy2021} coupling regimes. Other interesting on-chip circuits include split-ring resonators \cite{Wagle2024,Bhoi2017} or cross-line microwave circuits\cite{Wang2019}. In general coplanar waveguide (cpw) cavities produce strongly focused magnetic fields exhibiting pronounced non-homogeneities across the volume of nano-patterned magnets\cite{MartinezLosadelRincon2023, Hou2019, Li2019}.

Understanding the magnon-photon interactions in highly non-homogeneous ferromagnets and cavities is, therefore, of enormous technological and conceptual interest\cite{GonzalezGutierrez2024,Trif2024,Pan2024,Bondarenko2023,Psaroudaki2023,Sharma2022,Khan2021,Liensberger2021,Hrabec2020,MartinezPerez2019}. This issue has been addressed in Refs. \citenum{MartinezLosadelRincon2023} and \citenum{MartinezPerez2018} by combining electromagnetic and micromagnetic simulations. 
Here, we develop an alternative and user-friendly method to simulate the dynamics of magnon-polaritons of any type. To this end, we have modified the open-source micromagnetic code Mumax3\cite{Vansteenkiste2014}, harnessing its full potential for the simulation of magnetic systems. This includes several possible contributions to anisotropy such as shape, magnetocrystalline, or Dzyaloshinskii-Moriya interaction, and the possibility to introduce space- and time- dependent properties of ferromagnets. The new version is named Mumax3-cQED and is freely available on GitHub. Magnon-photon interaction is included through an effective retarded field that reflects the spatial- and time-dependence of the zero point magnetic field fluctuations created by the cavity. The latter can be simulated using other softwares (such as COMSOL or 3D-MLSI\cite{Khapaev2002}) and plugged into Mumax3-cQED. Along this paper we derive the new  equation of motion and show its implementation into Mumax3. As a benchmark, we use Mumax3-cQED to reproduce results from the Dicke model, a toy model that describes spin-photon coupling and can be solved analytically. Finally, we present a number of simulations that highlight the potential of Mumax3-cQED to reproduce experiments in which magnetic field or magnetization inhomogeneities play a fundamental role.


%% file: sections/01-llg-cavity.tex
\section{Landau-Lifshitz-Gilbert equation for a magnet coupled to a cavity}
\label{sec:llg-plus-cavity}

\subsection{The LLG equation in Mumax3}
\label{sec:llg}

Mumax3 computes the evolution of the reduced magnetization $\boldsymbol m_i$ (a unit vector) by numerically solving the LLG equation
\begin{equation}
    \dot{\boldsymbol m_i} \equiv \frac{\partial \boldsymbol m_i}{\partial t} = - \gamma \frac{1}{1 + \alpha^2} \left(\boldsymbol m_i \times \boldsymbol B_{\rm{eff}}(\boldsymbol r_i) + \alpha \boldsymbol m_i \times \left(\boldsymbol m_i \times \boldsymbol B_{\rm{eff}}(\boldsymbol r_i) \right) \right) \,,
    \label{eq:llg}
\end{equation}
with $\gamma > 0$ the gyromagnetic ratio, $\alpha$ a dimensionless damping parameter and $\boldsymbol B_{\rm{eff}}$ the effective field. Mumax3 performs a finite-difference discretization of space, dividing the ferromagnet into orthorombic cells and associating a reduced magnetization vector, $\boldsymbol m_i$, to the center of each cell. 
In Mumax3, the effective field can have the following contributions: external, $\boldsymbol B_{\rm{ext}}$, demagnetization, $\boldsymbol B_{\rm{demag}}$, exchange, $\boldsymbol B_{\rm{exch}}$, Dzyaloshinskii-Moriya, $\boldsymbol B_{\rm{dm}}$, magneto-crystalline anisotropy, $\boldsymbol B_{\rm{anis}}$ and thermal, $\boldsymbol B_{\rm{therm}}$ \cite{Vansteenkiste2014}. 
In the following, we will show that the effect of the cavity can be incorporated as an aditional contribution to $\boldsymbol B_{\rm{eff}}$: the cavity field, $\boldsymbol B_{\rm{cav}}$.

The first term on the right-hand side of Eq. \eqref{eq:llg} represents the precession of a spin around the effective magnetic field, $\boldsymbol{B}_{\rm{eff}}$, within a classical and mean-field approximation. This equation corresponds to the large spin limit of the Heisenberg equation for the spin operator, expressed as 
\begin{equation}
    \dot{\hat{\boldsymbol{S}_i}} = \frac{i}{\hbar} [\hat{H}, \hat{\boldsymbol{S}_i}] \,,
    \label{eq:heisenbergeomspin}
\end{equation}
where the spin operators satisfy 
\begin{equation}
    [\hat{S}_i^\alpha, \hat{S}_i^\beta] = i \hbar \epsilon_{\alpha \beta \gamma} \hat{S}_i^\gamma \,,
\end{equation}
with $\epsilon_{\alpha \beta \gamma}$ denoting the Levi-Civita symbol. Beware of the abuse of notation by using $i$ both as the imaginary unit and cell index in Eq. \eqref{eq:heisenbergeomspin} and hereinafter. In Eq. \eqref{eq:heisenbergeomspin}, $\hat {H}$ is the total Hamiltonian that accounts for all interactions and fields through $\boldsymbol{B}_{\rm{eff}}$. Within the mean-field treatment it is simply $\hat H = \gamma \sum_i\hat{\boldsymbol S_i} \cdot \boldsymbol B_{\rm eff}(\boldsymbol r_i)$. In the large spin limit, the spin operator can be approximated as a classical vector, $\boldsymbol{S}_i = \langle \hat{\boldsymbol{S}_i} \rangle$, with a magnitude of $\hbar S_i$. The magnetic dipole moment is defined as $\boldsymbol{\mu}_i = -\gamma \boldsymbol{S}_i$, and the magnetization as $\boldsymbol{M}_i = \boldsymbol{\mu}_i / V_{\rm{c}}$, where $V_{\rm{c}}$ is the cell volume. The reduced magnetization is then $\boldsymbol{m}_i = \boldsymbol{M}_i / M_{{\rm s}, i}$, with the saturation magnetization given by $M_{{\rm s}, i} = \hbar \gamma S_i / V_{\rm{c}}$. Consequently, the equation of motion for the magnetization is 
\begin{equation}
    \dot{\boldsymbol{m}_i} = - \gamma \boldsymbol{m}_i \times \boldsymbol{B}_{\rm{eff}}(\boldsymbol r_i) \,,
\end{equation}
consistent with the first term on the right-hand side of \eqref{eq:llg}.
The second term is the so-called phenomenological Gilbert damping. It induces a decay of the magnetization vector toward the precession axis determined by $\boldsymbol B_{\rm{eff}}$. 

We will focus on the modification of the first term by the cavity, and, once we establish how $\boldsymbol B_{\rm{eff}}$ is modified there, we will include this modification also in the $\boldsymbol B_{\rm{eff}}$ that appears on the second term, à la Gilbert.

\subsection{Introducing the effect of the cavity}
\label{sec:llgwithcavity}

We will consider a single mode cavity, in that case the Hamiltonian is \cite{RomanRoche2021, RomanRoche2022}
\begin{equation}
    \hat H = \gamma \sum_i \hat{\boldsymbol S_i} \cdot \boldsymbol B_{\rm{eff}}(\boldsymbol r_i) + \gamma \sum_i \hat{\boldsymbol S_i} \cdot \boldsymbol B_{\rm{rms}}(\boldsymbol r_i) \left(\hat a + \hat a^\dagger\right) + \hbar \omega_c \hat a^\dagger \hat a \,.
    \label{eq:startH}
\end{equation}
Here the spins are subject to both an effective field, $\boldsymbol B_{\rm{eff}}$, and the average cavity magnetic field $\boldsymbol B_{\rm{rms}}\left(\hat a + \hat a^\dagger\right)$, with $\hat a$, $\hat a^\dagger$ the bosonic annihilation and creation operators obeying cannonical commutation relations $[\hat a, \hat a^\dagger] = 1$ and $\omega_c$ the cavity frequency. The corresponding Heisenberg equations of motion are
\begin{align}
    & \dot{\hat{\boldsymbol S_i}}=-\gamma \hat{\boldsymbol S_i} \times \boldsymbol B_{\rm{eff}}(\boldsymbol r_i)-\gamma\left(\hat{\boldsymbol S_i} \times\boldsymbol B_{\rm{rms}}(\boldsymbol r_i)\right)\left(\hat a + \hat a^\dagger\right) \,, \label{eq:eqmotionspin} \\
    & \dot{\hat a} = -i \omega_{c} \hat a - i\frac{\gamma}{\hbar} \sum_i \hat{\boldsymbol S_i}\cdot \boldsymbol B_{\rm{rms}}(\boldsymbol r_i) \,. \label{eq:eqmotiona}
\end{align}
Hamiltonian \eqref{eq:startH} is essentially the Dicke model (See Sec. \ref{sec:benchmarking-Mumax3-CQED}). Its equations of motion [Eqs. \eqref{eq:eqmotionspin} and \eqref{eq:eqmotiona}] are fully quantum (within the mean-field approximation for spin-spin interactions). Consequently, the light-matter interaction generates spin-photon correlations. However, in the large $S$ limit, which is the focus here, these correlations can be disregarded \cite{carollo2021exactness}, consistent with the classical limit assumed for the spin dynamics in the LLG equation.
Armed with this information, if we take expected values in Eqs. \eqref{eq:eqmotionspin} and \eqref{eq:eqmotiona}, we can simplify
\begin{equation}
    \langle \left(\hat{\boldsymbol S_i} \times\boldsymbol B_{\rm{rms}}(\boldsymbol r_i)\right)\left(\hat a + \hat a^\dagger\right) \rangle \to \langle  \hat{\boldsymbol S_i} \rangle \times\boldsymbol B_{\rm{rms}} (\boldsymbol r_i) \langle \hat a + \hat a^\dagger\rangle \,,
\end{equation}
and arrive to classical equations of motion for the expected values of the spin and cavity degrees of freedom. This corresponds to a zeroth-order mean-field decoupling of light and matter that is exact if the initial states of the evolution are uncorrelated. The same mean-field decoupling at the level of the equations of motion but carried out to first order can be used to compute two-point correlators in cavity QED materials \cite{romanroche2024linear, romanroche2024cavity}. Defining $\alpha = \langle \hat a \rangle$ and $\alpha^* = \langle \hat a^\dagger \rangle$ we can write the equations of motion as
\begin{align}
    &\dot{\boldsymbol S_i}=-\gamma\boldsymbol S_i\times\boldsymbol B_{\rm{eff}}'(\boldsymbol r_i) \,, \label{eq:eqmotionspinclasical}\\
    &\dot \alpha = -i \omega_c \alpha - i \frac{\gamma}{\hbar} \sum_i \boldsymbol S_i \cdot \boldsymbol B_{\rm{rms}}(\boldsymbol r_i) \label{eq:eqmotionaclasical} \,,
\end{align}
with $\boldsymbol B_{\rm{eff}}'(\boldsymbol r_i) = \boldsymbol B_{\rm{eff}}(\boldsymbol r_i) + \boldsymbol B_{\rm rms}(\boldsymbol r_i)(\alpha + \alpha^*)$. As explained in Sec. \ref{sec:llg}, Equation \eqref{eq:eqmotionspinclasical} is the LLG equation with a modified effective field that depends on two additional cavity degrees of freedom. 
The system dynamics could be obtained by explicitly solving Eqs. \eqref{eq:eqmotionspinclasical} and \eqref{eq:eqmotionaclasical}. However, our goal is to incorporate the effect of the cavity while using the existing Mumax3 infraestructure, which is designed to handle only the magnetization degrees of freedom. For that reason, we will first integrate out the cavity by solving the associated equations of motion, to arrive to an effective description that depends only on the magnetization. This yields (See \ref{app:eom} for details)
\begin{equation}
    \dot{\boldsymbol m_i}=-\gamma\boldsymbol m_i\times\boldsymbol B_{\rm{eff}}'(\boldsymbol r_i) \,,
\end{equation}
with $\boldsymbol B_{\rm{eff}}'(\boldsymbol r_i) = \boldsymbol B_{\rm{eff}}(\boldsymbol r_i) + \boldsymbol B_{\rm{cav}}(\boldsymbol r_i)$ and
\begin{equation}
    \boldsymbol B_{\rm{cav}}(\boldsymbol r_i) = \boldsymbol B_{\rm{rms}}(\boldsymbol r_i) \Gamma(t) \,,
    \label{eq:bcav}
\end{equation}
where
\begin{equation}
\begin{multlined}
    \Gamma(t) =  \; 2 e^{- \kappa t} \Re \left(\alpha_0 e^{-i\omega_c t}\right) \\
    - \frac{2 V_{\rm c}}{\hbar} \int_0^t d\tau e^{ \kappa (\tau - t)} \sin(\omega_c (\tau - t)) \sum_i M_{{\rm s}, i}  \boldsymbol m_i(\tau) \cdot  \boldsymbol B_{\rm{rms}}(\boldsymbol r_i)\,.
\end{multlined}
    \label{eq:gammafinal}
\end{equation}
With this, the effect of the cavity can be incorporated in Mumax3 as a new contribution to the effective field: $\boldsymbol B_{\rm{eff}} \to \boldsymbol B_{\rm{eff}}' = \boldsymbol B_{\rm{eff}} + \boldsymbol B_{\rm{cav}}$. This new cavity field, $\boldsymbol B_{\rm{cav}}$ \eqref{eq:bcav}, is the average magnetic field of the cavity, $\boldsymbol B_{\rm{rms}}$, times a memory factor, $\Gamma$. In turn, the differential equation that is the LLG equation has become an integro-differential equation (IDE) by the appearance of the memory term $\Gamma(t)$, which depends on the full magnetization history. We can understand this as the mathematical reflection of the fact that we have eliminated the cavity from our dynamical description and, as a consequence, a cavity-mediated retarded interaction between the spins appears. Much like dissipation is introduced phenomenologically in the LLG equation with the Gilbert term, we have introduced dissipation in the cavity by promoting $\omega_c \to \omega_c - i \kappa$ in Eq. \eqref{eq:eqmotiona}. This corresponds to local dissipation at the level of a Linblad master equation for the cavity.

Despite having eliminated $\alpha$ and $\alpha^*$ as dynamical variables, one might be interested in their values. For instance, $|\alpha|^2 = \langle a^\dagger a \rangle$ is the number of photons in the cavity. These can be computed a posteriori from the magnetization, $\boldsymbol m(t)$, obtained from a Mumax3-cQED simulation. They are given by (See \ref{app:eom} for details)
\begin{equation}
    \alpha = \alpha_0  e^{-\kappa t} e^{-i\omega_c t} + i\frac{V_{\rm{c}}}{\hbar} \int_0^t d\tau e^{\kappa (\tau - t)} e^{i\omega_c(\tau-t)} \sum_i M_{{\rm s}, i}  \boldsymbol m_i(\tau)  \cdot \boldsymbol B_{\rm{rms}}(\boldsymbol r_i)  \,, \\
\end{equation}

%% file: sections/02-mumax-modification.tex
\section{From Mumax3 to Mumax3-cQED}
\label{sec:mumax3-modification}

\subsection{Adapting the memory term for efficient computation}
\label{sec:recursive memory term}

In this section we explain how we have implemented the cavity contribution to the effective field, $\boldsymbol B_{\rm{cav}}$ \eqref{eq:bcav}, into Mumax3-cQED without making any modification to the existing ordinary differential equation (ODE) integrator in Mumax3. First, it is convenient to decouple the time dependence on $t$, the current time, and $\tau$, the past time, in the integral of Eq. \eqref{eq:gammafinal}, such that
\begin{equation}
     \Gamma(t) = 2 e^{- \kappa t} \Re \left(\alpha_0 e^{-i\omega_{\rm c} t}\right) - \frac{2 V_{\rm c}}{\hbar}  e^{-\kappa t} \left( \cos(\omega_{\rm c} t) S(t) - \sin(\omega_{\rm c} t) C(t)\right) \,,
\end{equation}
with
\begin{align}
    & S(t) = \int_0^t d\tau e^{\kappa \tau} \sin(\omega_{\rm c} \tau) \sum_i M_{{\rm s}, i} \boldsymbol m_i(\tau) \cdot \boldsymbol B_{\rm{rms}}(\boldsymbol r_i)  \,, \\
    & C(t) = \int_0^t d\tau e^{\kappa \tau} \cos(\omega_{\rm c} \tau) \sum_i M_{{\rm s}, i} \boldsymbol m_i(\tau) \cdot \boldsymbol B_{\rm{rms}}(\boldsymbol r_i)  \,.
\end{align}
Note that here the integrands no longer depend on the current time, $t$. So far we have dealt with time as a continuous variable, but in Mumax3, time is discretized. Expressing $S(t)$ and $C(t)$ in terms of discrete time, $t \to t_n = n \Delta t$ yields
\begin{align}
    & S_n = S(t_n) = S_{n-1} + e^{\kappa t_n} \sin(\omega_{\rm c} t_n) \sum_i M_{{\rm s}, i} \boldsymbol m_i(t_n) \cdot \boldsymbol B_{\rm{rms}}(\boldsymbol r_i) \Delta t \,,  \label{eq:Sdiscrete} \\
    & C_n = C(t_n) = C_{n-1} + e^{\kappa t_n} \cos(\omega_{\rm c} t_n) \sum_i M_{{\rm s}, i} \boldsymbol m_i(t_n) \cdot \boldsymbol B_{\rm{rms}}(\boldsymbol r_i) \Delta t \,. \label{eq:Cdiscrete}
\end{align}
with $S_0 = C_0 = 0$.
This recursive definition is the key to efficiently implementing $\boldsymbol B_{\rm{cav}}$ within Mumax3. By storing $S_n$ and $C_n$ as persistent variables across consecutive calls to the integrator we avoid storing the full history of magnetizations, saving memory, and avoid resumming the full history at each time step, optimizing run speed.

Finally, to avoid handling complex variables in Mumax3 we express the term $\Re \left(\alpha_0 e^{-i\omega_{\rm c} t}\right)$ in terms of the real and imaginary parts of $\alpha_0$, which yields
\begin{equation}
    \Gamma(t_n) = e^{- \kappa t_n} \cos (\omega_{\rm c} t_n) \left(x_0 - \frac{2 V_{\rm c}}{\hbar} S_n\right) - e^{- \kappa t_n} \sin (\omega_{\rm c} t_n) \left(p_0 - \frac{2 V_{\rm c}}{\hbar} C_n\right) \,,
    \label{eq:gammadiscretefinal}
\end{equation}
with $x_0 = \alpha_0 + \alpha_0^* = 2 \Re (\alpha)$ and $p_0 = \alpha_0 - \alpha_0^* = - 2 \Im (\alpha)$.


\subsection{Workflow in Mumax3-cQED}
\label{subsec:workflow-new-code}

Mumax3-cQED has been implemented as a fork of Mumax3\footnote{\url{https://github.com/mumax/3}}. 
The source code, including detailed installation intructions for UNIX and Windows systems, is hosted in a Git repository\footnote{\url{https://github.com/Mumax3-cQED/mumax3-cqed}}. A summary of the source code changes implemented in Mumax3-cQED is provided in \ref{app:source}.
The new Mumax3-cQED source code has been tested in Windows 10 with \italic{go 1.9} and \italic{CUDA 10.2} and in \italic{Debian 12 Bookworm} with \italic{go 1.9} and \italic{CUDA 12.0}. In both cases a GPU \italic{NVIDIA RTX A4000} has been used.

After installation, Mumax3-cQED follows a script-based workflow inherited from Mumax3. 
Mumax3-cQED accepts an input script that defines the parameters and initial conditions of the micromagnetic simulation and returns output data representing the results of the simulation. We have introduced a new set of built-in instructions for the scripting language that extend the existing API of Mumax3:
\begin{itemize}

\item \lstinline{B_rms} (type \verb|vector|): Zero-point magnetic field (vacuum field) of the cavity (See Eq. \eqref{eq:startH}). This is a three-component vector, it can take a single homogeneous value or it can take spatially-dependent values loaded from an OVF file (units \verb|T|).

\item \lstinline|Wc| (type \verb|float64|): Resonant frequency of the cavity (See Eq. \eqref{eq:startH}) (units \verb|rad| $\cdot$ \verb|s|\super{-1}).

\item \lstinline|Kappa| (type \verb|float64|): Cavity dissipation rate (See Eq. \eqref{eq:gammafinal}) (units \verb|rad| $\cdot$ \verb|s|\super{-1}).

\item \lstinline|X0| (type \verb|float64|): Initial value of $x_0 = 2 \Re (\alpha)$ (See Eq. \eqref{eq:gammadiscretefinal}) (default value 0).

\item \lstinline|P0| (type \verb|float64|): Initial value of $p_0 = -2 \Im (\alpha)$ (See Eq. \eqref{eq:gammadiscretefinal}) (default value 0).

\item \lstinline|HBAR| (type \verb|float64|): Reduced Planck constant (default value $1.05457182 \mathrm{e}{-34}$ \verb|J| $\cdot$ \verb|s|).

\item \lstinline|ResetMemoryTerm()| (type \verb|func|): Function to reset the memory term $\Gamma$ (See Eq. \eqref{eq:gammadiscretefinal}). Resetting the memory term amounts to establishing the current state of the simulation as corresponding to the initial time, $t = 0$, at which the cavity and the magnet couple. This allows to chain different simulations in the same script.

\item \lstinline|CavityFeatureStatus| (type \verb|int|): Read-only variable to check whether the cavity feature is active or not, it can be invoked from the print as \verb|print(CavityFeatureStatus)|. It offers a sanity check, it returns 1 if \lstinline{B_rms} has been set (cavity enabled) and 0 otherwise (cavity disabled).

\end{itemize}

In \ref{app:brms} we discuss how the cavity $\boldsymbol B_{\rm rms}$ can be computed with other software and in \ref{app:micromag} we provide an example script showing how it can be loaded into Mumax3-cQED and used in a simulation. 
In addition to these variables that set the cavity parameters, we have also introduced some control instructions to provide the user with detailed information about the simulation:
\begin{itemize}

\item \lstinline|ShowSimulationSummary| (type \verb|boolean|): Whether to show a summary of the simulation after calling the \verb|run()| function in the script. The summary is printed to the console and to the log file (default value \verb|true|).

\item \lstinline|PrintScriptExecutionTime()| (type \verb|func|): Function to time the script execution or a portion of it, depending on the value of \lstinline|StartCheckPoint|. It prints the start and end times and their difference to the console and to the log file. 

\item \lstinline|StartCheckPoint| (type \verb|date|): Variable to set the start time for the timing of the script execution. In principle, it must be set to \verb|StartCheckPoint = now()| anywhere in the code to time the script execution from that point onwards (default value \verb|now()| called at the beginning of the script execution).

\end{itemize}
%

%% file: sections/03-benchmarking-mumax-cqed.tex
 \section{Benchmark}
\label{sec:benchmarking-Mumax3-CQED}

Our testbed will be the Dicke model. Although originally conceived to describe electric dipoles coupled to a cavity\cite{Dicke1954}, it can also be applied to magnetic dipoles, i.e. spins \cite{RomanRoche2021}. We choose the Dicke model for several reasons. First, it can be implemented in Mumax3-cQED using one single cell, i.e., it does not benefit from the GPU-reliant parallelization. Therefore, its evolution can also be numerically simulated in Python (or any other programming language) with reasonable CPU runtimes. This is done by simply integrating the joint LLG equation and equations of motion for the cavity degrees of freedom associated to Eqs. \eqref{eq:eqmotionspinclasical} and \eqref{eq:eqmotionaclasical} with the default integrators included in Python's Scipy. This unsophisticated implementation in Python provides the transparency that the highly optimized Mumax3 code lacks. It also allows us to check to what degree the implicit integration (with the magnetization as sole degree of freedom) that we implement in Mumax3-cQED is equivalent to an explicit integration of the original equations of motion. Second, the Dicke model can be solved exactly at equilibrium \cite{Hepp1973, wang1973phase}, providing values of observables, like the magnetization, to which we can expect the dynamics to converge at large times. It also gives the frequencies of the polaritons that constitute the normal modes of this hybrid system \cite{Emary2003}, which can be compared with the Fourier spectrum of the time evolution obtained with Mumax3.

The ground-state sector of the Dicke Hamiltonian is described by \cite{kirton2018introduction}
\begin{equation}
    \hat H  = \omega_z \hat S_z + \hbar \omega_c \hat a^\dagger \hat a + \lambda \sqrt{\frac{2}{S}}(\hat a + \hat a^\dagger) \hat S_x \,
    \label{eq:dickeH}
\end{equation}
with $\hat S_\alpha$ spin $S$ operators obeying $[\hat S_\alpha, \hat S_\beta] = i \hbar \epsilon_{\alpha \beta \gamma} \hat S_\gamma$. These spin $S$ operators are typically understood as maximum total spin operators of an underlying system of $N$ spin $1/2$ particles, with $S = N/2$. Here, $\omega_z$ is the bare-spin energy splitting, $\omega_{\rm c}$ is the cavity frequency and $\lambda$ is the collective light-matter coupling. By comparing Eq. \eqref{eq:dickeH} with Eq. \eqref{eq:startH} we see that we can implement the Dicke model in Mumax3-cQED by identifying
\begin{align}
    & \boldsymbol B_{\rm{eff}} = \boldsymbol B_{\rm{ext}} = \left(0, 0, \frac{\omega_z}{\gamma} \right) \,, \\
    & \boldsymbol B_{\rm{rms }} = \left(\sqrt{\frac{2}{S}}\frac{\lambda}{\gamma}, 0, 0 \right) \,.
\end{align}
This corresponds to a single cell, i.e. a single magnetization vector, under the effect of an external field and the cavity field: $\boldsymbol B_{\rm eff} = \boldsymbol B_{\rm ext} + \boldsymbol B_{\rm cav}$.

The Dicke model can be solved exactly at equilibrium in the thermodynamic limit, $S, N \to \infty$. At zero temperature the magnetization along $x$ is \cite{wang1973phase, romanroche2023exact}
\begin{align}
    & m_x = \begin{cases}
        0  & \rm{if} \quad \lambda < \lambda_{\rm c} \,, \\
        \pm \sqrt{1 - \mu^2}  & \rm{if} \quad \lambda > \lambda_{\rm c} \,,
    \end{cases}
    \label{eq:dickeeqmag}
\end{align}
with $m_x = \langle S_x \rangle / (\hbar S)$, $\mu = (\lambda_c / \lambda)^2$ and $\lambda_{\rm c} = \sqrt{\omega_{\rm c} \omega_z} / 2$. The model exhibits a phase transition between a normal paramagnetic phase with $m_x = 0$ and a superradiant ferromagnetic phase with $m_x \neq 0$. The symmetry between the positive and negative values of $m_x$ is spontaneously broken at the phase transition. The Dicke model can be cast to a two-oscillator model with a Holstein-Primakov transformation of the spin operators. The resulting quadratic model can be solved exactly in each phase to yield the frequencies of the normal modes of the system, the polaritons \cite{Emary2003}
\begin{equation}
    2 \Omega_\pm^2 = \begin{cases}
        \omega_z^2 + \omega_{\rm c}^2 \pm \sqrt{\left(\omega_z^2 - \omega_{\rm c}^2\right)^2 + 16 \lambda^2 \omega_z \omega_{\rm c}} & \rm{if} \quad \lambda < \lambda_{\rm c} \,, \\
        \omega_z^2/\mu^2 + \omega_{\rm c}^2 \pm \sqrt{\left(\omega_z^2/\mu^2 + \omega_{\rm c}^2\right)^2 + 4 \omega_z^2 \omega_{\rm c}^2} & \rm{if} \quad \lambda > \lambda_{\rm c} \,,
    \end{cases}
    \label{eq:dickepolaritons}
\end{equation}
Equations \eqref{eq:dickeeqmag} and \eqref{eq:dickepolaritons} provide reference values to benchmark the simulation of the Dicke model with Mumax3-cQED.

\begin{figure}
    \centering
    \includegraphics[width =\columnwidth]{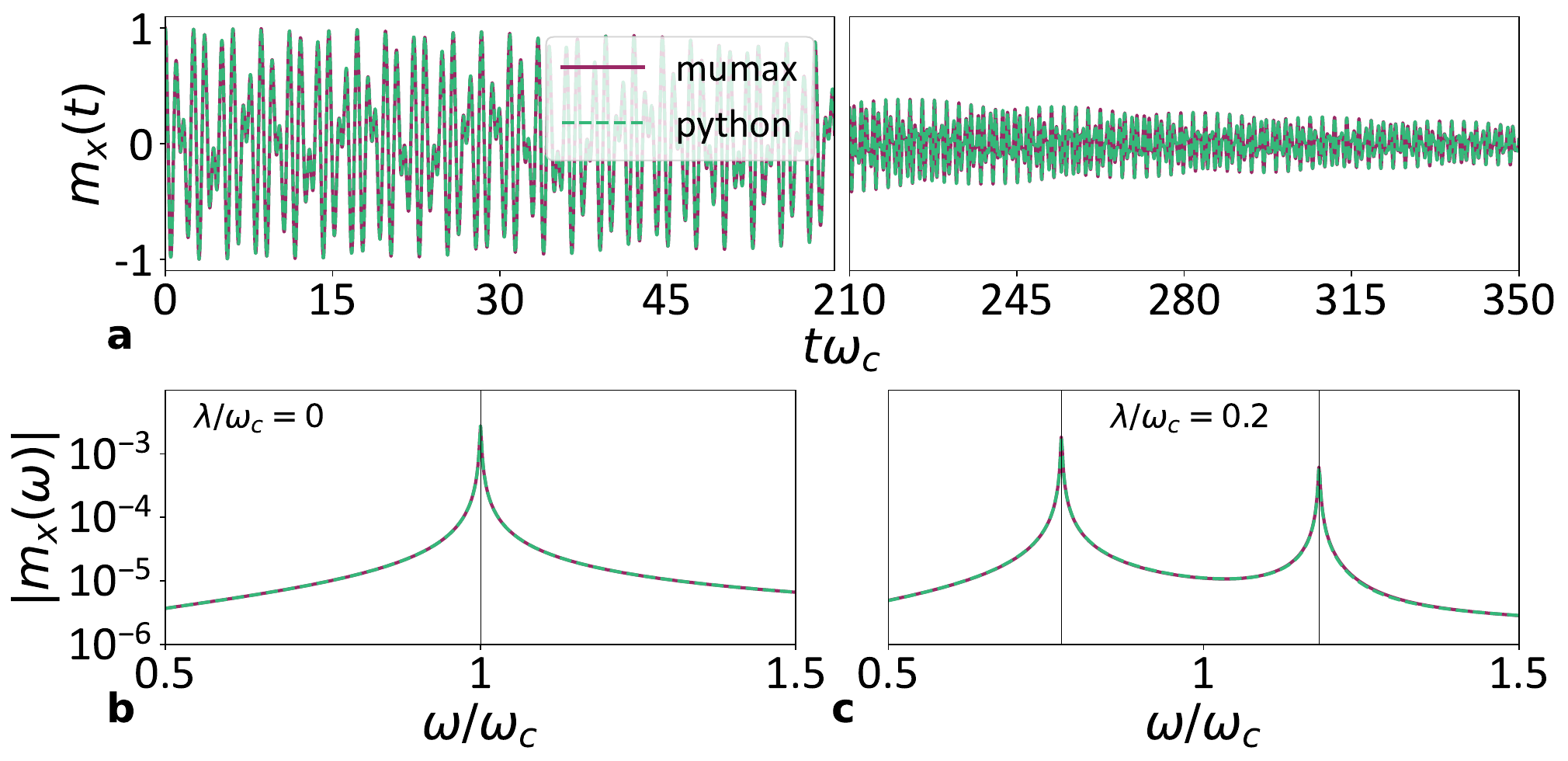}
    \caption{Time evolution of the Dicke model in the normal paramagnetic phase, $\lambda < \lambda_{\rm c}$, for $\omega_z = \omega_c$. a: Magnetization as a function of time, for small and large times. b and c: Fourier spectra of the magnetization for zero (b) and non-zero (c) coupling. The vertical black lines mark the frequencies of the polaritons computed analytically.}
    \label{fig:dickenormal}
\end{figure}
Fig. \ref{fig:dickenormal}(a) shows an example of a time evolution of $m_x$ computed with Mumax3-cQED within the normal paramagnetic phase, $\lambda < \lambda_{\rm c}$. It matches the evolution computed with Python. The magnetization converges to zero at large times, consistent with the equilibrium value [Cf. Eq. \eqref{eq:dickeeqmag}]. Figs. \ref{fig:dickenormal}(b) and (c) show the Fourier transform of two time evolutions, like the one shown in Fig. \ref{fig:dickenormal}(a). The Fourier spectra present peaks at the polaritonic frequencies of the Dicke model,  at $\omega = \omega_z = \omega_c$ for $\lambda = 0$ and at $\omega = \Omega_\pm$ \eqref{eq:dickepolaritons} for non-zero $\lambda$. 
\begin{figure}
    \centering
    \includegraphics[width = \columnwidth]{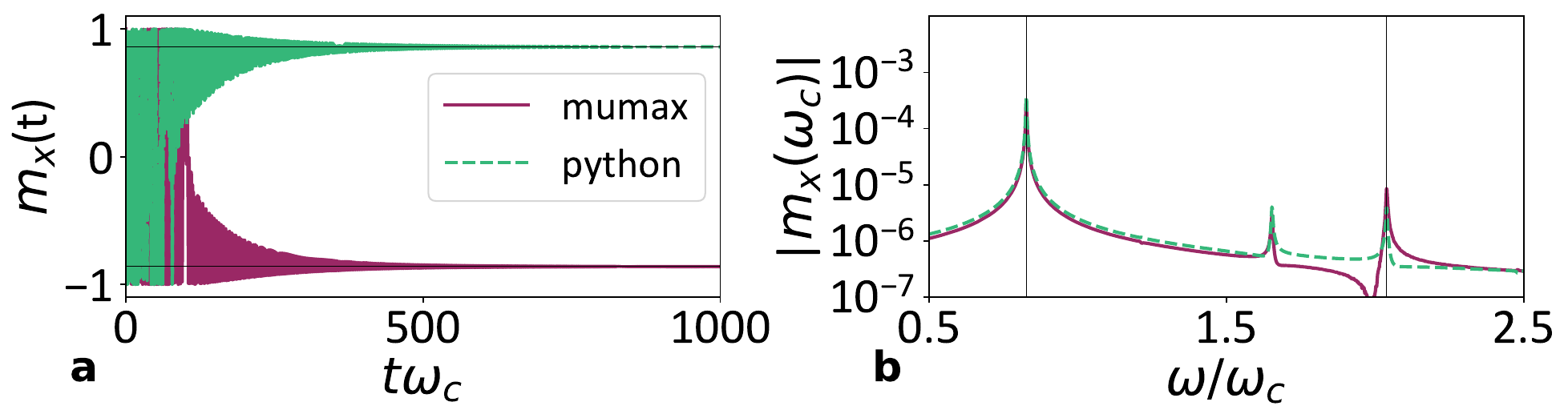}
    \caption{Time evolution of the Dicke model in the superradiant ferromagnetic phase, $\lambda/\lambda_{\rm c} = 1.4$, for $\omega_z = \omega_c$. a: Magnetization as a function of time. The horizontal black lines mark the possible equilibrium values of the magnetization computed analytically. b:  Corresponding Fourier spectra of the magnetization. The vertical black lines mark the frequencies of the polaritons computed analytically.}
    \label{fig:dickesuper}
\end{figure}

The dynamics within the superradiant ferromagnetic phase serve to explicitly demonstrate the numerical differences between the IDE integration in Mumax3-cQED and the ODE approach in our custom Python code. In the broken-symmetry phase, the two possible equilibrium solutions, $\pm \sqrt{1 - \mu^2}$ \eqref{eq:dickeeqmag}, act as attractors or fixed points of the system. Consequently, even minor differences may lead the system to converge to different attractors. However, this divergence has no physical consequences. As shown in Figure \ref{fig:dickesuper}, numerical variations result in the system reaching distinct attractors, but both solutions converge to the correct equilibrium values. This consistency emphasizes that dissipation is introduced in a thermodynamically consistent manner.

We have conducted tests analogous to the ones presented in Figs. \ref{fig:dickenormal} and \ref{fig:dickesuper} in the full range of values of $\omega_c$, $\omega_z$ and $\lambda$ with similar results. The differences between the Mumax3-cQED and Python evolutions grow progresively with increasing coupling $\lambda$ and are only noticeable well into the ultra-strong coupling regime, as shown in Fig. \ref{fig:dickesuper}. In any case, Mumax3-cQED always produces one of the correct equilibrium values for the different observables and Fourier spectra peaked at the correct polaritonic frequencies.  We take this as confirmation that Mumax3-cQED works as intended. It correctly captures the hybridization of light and spins, even in non-trivial ordered phases and ultra-strong coupling regimes and it enables the study of non-equilibrium dynamics in magnonic QED.

%% file: sections/04-nanowire-cavity.tex
\section{Ferromagnetic materials in a cavity}
\label{sec:ferromagnetic-materials-cavity}

We finally show the potential of Mumax3-cQED applied to two relevant non-analytical problems. The first focuses on a saturated ferromagnet coupled to a realistic resonator which generates strongly non-homogeneous magnetic field modes. In the second example, we will calculate the response of a spin texture interacting with cavity photons.

\subsection{Effects of the parity of the magnetic mode}
\begin{figure}
    \centering
    \includegraphics[width =0.9\columnwidth]{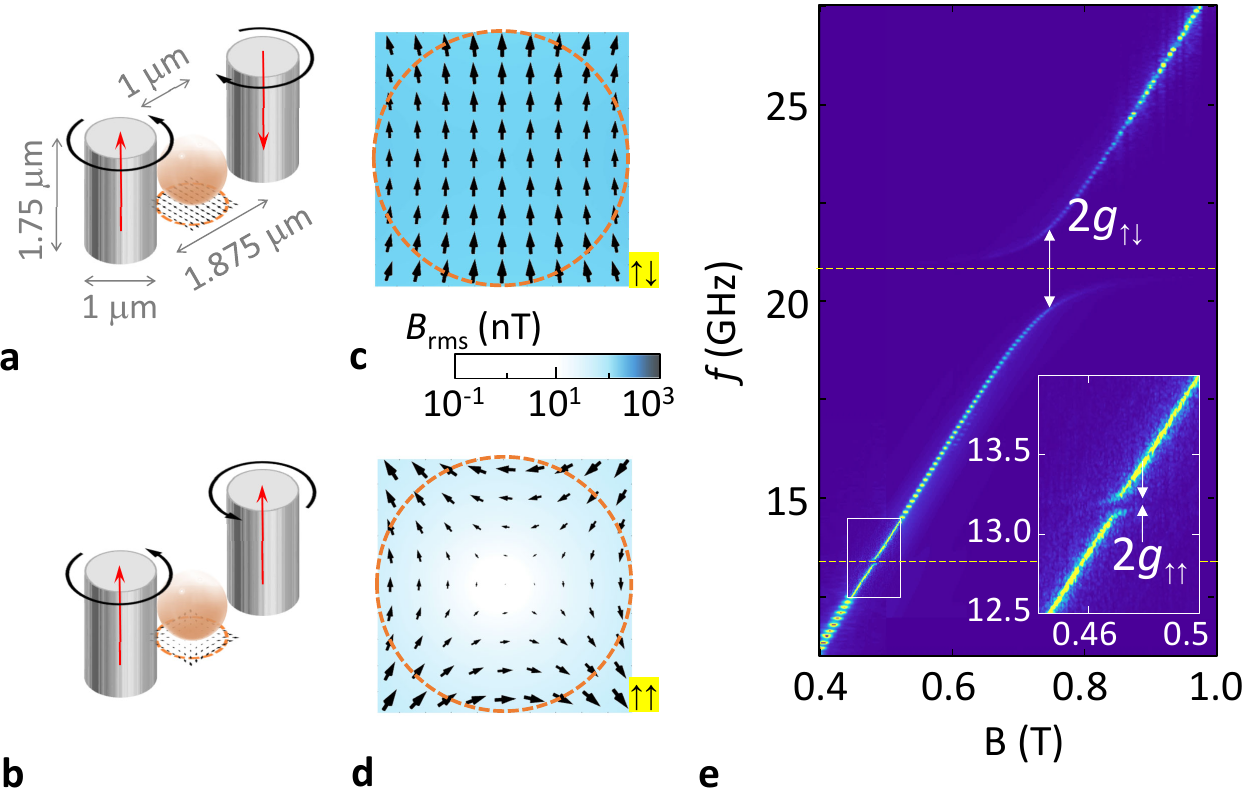}
    \caption{Two-post re-entrant cavity with electric currents (red arrows) flowing anti-parallel (a: bright mode) or parallel (b: dark mode). Resulting magnetic field lines are represented by black arrows. Spatial distribution of $B_{\rm{rms}}$ for the bright (c) and dark (d) modes with the YIG sphere position highlighted in dashed orange. e: Response of the YIG sphere (inset: region of coupling to the dark mode. Dashed yellow lines highlight the frequencies of the bright and dark modes.}
    \label{fig:Fig3}
\end{figure}

We first consider a three-dimensional cavity with a "re-entrant" element, i.e., a metallic post or rod that extends into the cavity from one side. This element modifies the electromagnetic field distribution, concentrating the magnetic field near the bottom part of the post. On the other hand, an antinode of the electric field is created at the gap, between the post and the top of the cavity\cite{McAllister2017}. Using two posts results in two non-degenerate modes with anti-parallel and parallel currents. These currents produce even and odd field distributions between the two posts, referred to as bright and dark modes, respectively (see Fig. \ref{fig:Fig3}a and b). Here, we will reproduce the results obtained by Goryachev et al.\cite{Goryachev2014} using an yttrium iron garnet (YIG) sphere (radius of 0.4 mm) located between the two posts of a re-entrant cavity (internal radius of 5 mm, height 1.4 mm, post radius of 0.4 mm, post gap of 73 $\mu$m, and distance between the posts 1.5 mm) with dark and bright modes at frequencies $f_{\uparrow\downarrow} = 20.8$ GHz and $f_{\uparrow\uparrow} = 13.2$ GHz, respectively.

To obtain reasonable simulation times we scale down the whole system by a factor of 800 resulting into a YIG sphere with radius 500 nm ($\gg \lambda_{ex} \sim 14 $ nm, the exchange length of YIG). As demonstrated in Refs. \citenum{MartinezLosadelRincon2023} and \citenum{MartinezPerez2018}, the coupling strength is $g \propto B_{\rm{rms}} \sqrt{V_m}$, with $V_m$ the volume of the magnet. Therefore, to compensate the reduced volume of the system, and yield a coupling $g$ equivalent to the experiment, we also need to scale up $B_{\rm{rms}}$ by a factor of $\sim (800)^{3/2}$. The spatial-dependence of $\boldsymbol B_{\rm{rms}}(\boldsymbol r_i)$ created by the bright and dark modes are simulated in COMSOL as described in \ref{app:brms}.  For the bright mode, two anti-parallel currents yield an even, quasi homogeneous distribution of zero-point magnetic field fluctuations (see Fig. \ref{fig:Fig3}c). 
For the dark mode, two parallel identical currents result in an odd, highly non-homogeneous distribution (see Fig. \ref{fig:Fig3}d). 

Polariton dynamics are then calculated using Mumax3-cQED (see  \ref{app:micromag}) and shown in Fig. \ref{fig:Fig3}e. Due to symmetry reasons, the bright mode can excite and couples strongly to the Kittel mode in the YIG sphere, yielding a coupling energy of $g_{\uparrow\downarrow}/2\pi \sim 1$ GHz, as reported by Goryachev et al. On the other hand, the odd-symmetric dark mode couples very weakly to the Kittel precession, leading to a much weaker $g_{\uparrow\uparrow}/2\pi \sim 30$ MHz. We note that the excitation of additional (experimentally visible) magnetostatic modes cannot be reproduced. These modes are neither visible in simulations performed with the unmodified Mumax3, probably due to the strong reduction of the sphere radius (factor of $800$). 

In its current state, Mumax3-cQED considers a single-mode cavity field. However, our simulations demonstrate that results for multimode cavities can also be reproduced by simulating each mode individually. This is true as long as the cavity frequencies are well-separated, at least as much as the coupling strength. In any case, extending Mumax3-cQED to handle multimode cavities is straightforward and will be included in future releases.

\subsection{Spin textures}
\label{sec:spintextures}

Finally, we will simulate magnon-polariton dynamics in magnetic vortices. Spin textures such as domain walls, vortices and skirmions are topological solitons that can be created, annihilated and displaced with minimal distortion, offering attractive possibilities for classical and quantum information processing\cite{Yu2021}. Quantum states encoded in spin textures can be initialized, entangled and read out using superconducting or magnonic cavities \cite{Trif2024,Psaroudaki2023,Pan2024,Khan2021,Liensberger2021}. Vortices are also appealing for implementing nanoscopic cavities, with the postential to readout individual spins\cite{GonzalezGutierrez2024} or to mediate strong interactions between spin qubits and superconducting microcircuits\cite{MartinezPerez2018}. Magnetic vortices are easily stabilized in mesoscopic magntetic thin-film discs with lateral sizes between a few $100$ nm up to several $\mu$m \cite{Shinjo2000,Wachowiak2002}. Minimization of the magnetostatic energy yields an in-plane spiral distribution of spins with an out-of-plane vortex core (see Fig. \ref{fig:Fig1}b). Cylindrical symmetry yields a number of magnetostatic (azimuthal and radial) modes whereas the central soliton-like vortex exhibits a peculiar translation mode \cite{Guslienko2006}. This is the distinctive mode in which the vortex core gyrates around its equilibrium position. The sense of gyration is determined solely by the vortex's polarity, while the frequency is determined by the radius-to-thickness ratio. The experimental observation of strong coupling between cavity photons and the gyrotropic mode in magnetic vortices is still awaiting.

\begin{figure}
    \centering
    \includegraphics[width =0.9\columnwidth]{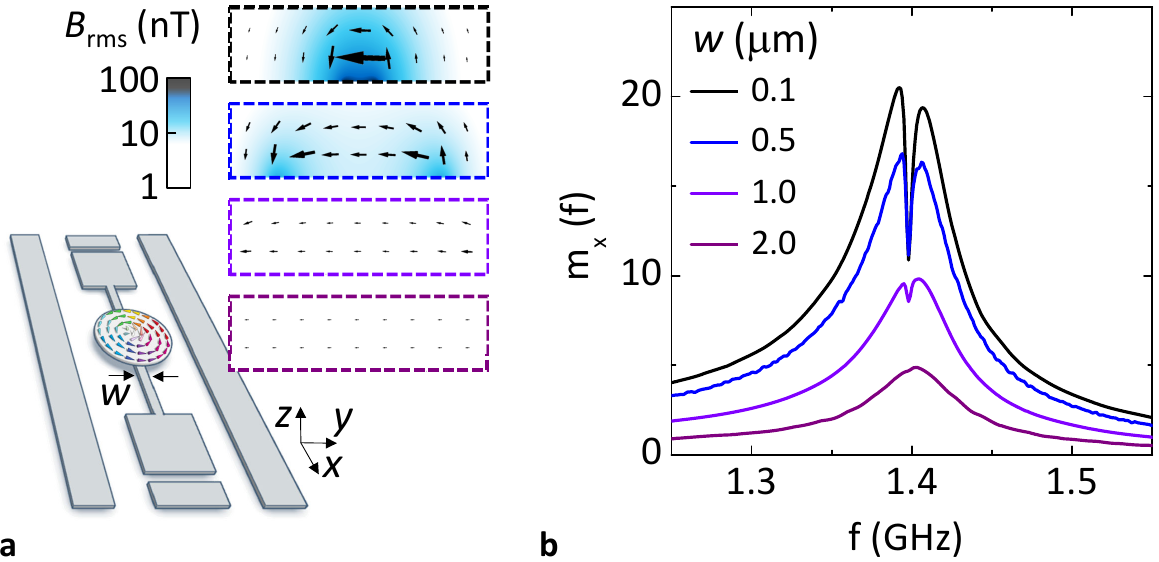}
    \caption{a: Py disc located at the magnetic antinode of a cpw distributed resonator of width $w$. The non-homogeneous distribution of zero-point field fluctuations at the disc position ($yz$ plane, dashed rectangles) is shown for different values of $w$ (from top to bottom, $0.1$, $0.5$, $1$ and $2$ $\mu$m).  b: Response of the Py disc vs. frequency for different values of $w$ (in $\mu$m). Decreasing $w$ allows going from the weak to the strong coupling regime.}
    \label{fig:Fig5}
\end{figure}

We chose a Py disc of radius $400$ nm and thickness $150$ nm in which a magnetic vortex is naturally stabilized  resulting in a gyrotropic frequency of $f=1.4$ GHz.  The disc is coupled to the magnetic antinode of a cpw resonator, with central conductor cross-section of thickness $50$ nm and width $w$. The corresponding distribution of $\boldsymbol B_{\rm{rms}}(\boldsymbol r_i)$ is calculated as described in \ref{app:brms} and shown in the inset of Fig. \ref{fig:Fig5}a. Decreasing $w$ has the effect of reducing the mode volume of the cavity. Consequently, the intensity of $B_{\rm rms}$ increases for the smallest linewidth of the central transmission line (upper panel in Fig. \ref{fig:Fig5}a). Fig. \ref{fig:Fig5}b shows the response of the Py disc in frequency domain calculated using Mumax3-cQED (see  \ref{app:micromag}). Decreasing $w$ allows reaching the strong vortex-photon regime as demonstrated by the opening of an anti-crossing for the minimum $w=100$ nm, yielding a splitting of $2 g/2 \pi=15$ MHz.

\section{Conclusions}
\label{sec:conclusions}

Mumax3-cQED is an open-source micromagnetic package derived from Mumax3. Besides modeling standard spin-spin interactions (e.g., dipolar, exchange, Dzyaloshinskii-Moriya etc) and coupling of magnetic moments to external magnetic fields, Mumax3-cQED incorporates the interaction of spins with cavity photons. This unique feature allows addressing magnon-
light interactions across all coupling
regimes, i.e., weak, strong, and even the ultra-strong
coupling regimes, combined with the GPU-accelerated potential of Mumax3. 
Dissipation is also included, allowing for the study of non-equilibrium dynamics and the approach to equilibrium.
The code has been verified by applying it to the case of a macrospin coupled to a cavity. This example is well-described by the Dicke model, which can be solved numerically and analytically. Furthermore, the software has been able to reproduce the results of the Dicke model even in the superradiant regime. Finally, we have demonstrated the  potential of Mumax3-cQED applied to various relevant examples in the field of cavity magnonics. First, a re-entrant two-post cavity where we observed the selective coupling of the Kittel mode to even or odd field distributions. Finally, we analyzed the dynamics of a magnetic vortex, a highly non-homogeneous texture,  in the regime ranging from weak to strong coupling.

Mumax3-cQED can now be used by the scientific community to design magnon-cavity experiments under two conditions: that the dynamics of the ferromagnet can be reliably reproduced with Mumax3, and that the spatial distribution of magnetic fields created by the cavity can be simulated (using another software). Mumax3-cQED allows increasing the power of simulations in cavity magnonics to the next level. With this code, it will be possible to study the influence of the cavity design on magnon-photon couplings and bring spin textures into play. The latter include  domain walls\cite{Trif2024}, vortices \cite{GonzalezGutierrez2024,Bondarenko2023,MartinezPerez2019} and skyrmions\cite{Pan2024,Psaroudaki2023,Khan2021,Liensberger2021,Hrabec2020}, attractive for applications in quantum devices and quantum sensing.

\section{Software availability}
The software used in this paper is called Mumax3-cQED and is available at \url{https://github.com/Mumax3-cQED/mumax3-cqed}
\section{CRediT authorship contribution statement}
\textbf{Sergio Mart\'inez-Losa del Rinc\'on}: Methodology, Software, Validation, Investigation, Formal analysis, Writing - Original Draft, Visualization
\textbf{Juan Rom\'an-Roche}: Methodology, Software, Validation, Investigation, Formal analysis, Writing - Original Draft, Visualization
\textbf{Andr\'es Mart\'in-Megino}: Methodology, Investigation, Formal analysis
\textbf{David Zueco}: Conceptualization, Methodology, Writing - Review \& Editing, Supervision, Funding acquisition
\textbf{Mar\'ia Jos\'e Mart\'inez-P\'erez}: Conceptualization, Methodology, Validation, Investigation, Formal analysis, Writing - Review \& Editing, Visualization, Supervision, Funding acquisition

\section{Declaration of Competing Interest}
The authors declare that they have no known competing financial interests or personal relationships that could have appeared to influence the work reported in this paper.

\section{Acknowledgments}

This work is partly funded and supported by the European Research Council (ERC) under the European Union’s Horizon 2020 research and innovation programme (948986 QFaST), the Spanish MCIN/AEI/10.13039/501100011033, the European Union NextGenerationEU/PRTR and FEDER through projects PID2022-140923NB-C21 and TED2021-131447B-C21, the CSIC program for the Spanish Recovery, Transformation and Resilience Plan funded by the Recovery and Resilience Facility of the European Union, established by the Regulation (EU) 2020/2094, the CSIC Research Platform on Quantum Technologies PTI-001, the Arag\'{o}n Regional Government through project QMAD (E09\_23R) and MCIN with funding from European Union NextGenerationEU (PRTR-C17.I1) promoted by the Government of Arag\'{o}n. J. R-R acknowledges support from the Ministry of Universities of the Spanish Government through the grant FPU2020-07231.

%% file: sections/appendix_i.tex
\section{Solving the equation of motion of the bosonic degrees of freedom}
\label{app:eom}

We use a change of variables to a ``rotating frame'': $\bar \alpha = \alpha e^{i \omega_c t}$, $\bar{\boldsymbol S} = \boldsymbol S e^{i \omega_c t}$. Using these rotating variables, Eq. \eqref{eq:eqmotiona} becomes
\begin{equation}
    \dot{\bar \alpha} = - i \frac{\gamma}{\hbar} \sum_i\bar{\boldsymbol S_i} \boldsymbol B_{\rm{rms}}(\boldsymbol r_i) \,,
\end{equation}
which is solved by
\begin{equation}
    \bar \alpha = \alpha_0 - i\frac{\gamma}{\hbar} \int_0^t d\tau \sum_i \bar{\boldsymbol S_i}(\tau) \cdot \boldsymbol B_{\rm{rms}}(\boldsymbol r_i)  \,.
\end{equation}
Here, $\alpha_0 = \bar \alpha_0 = \alpha (t = 0)$. Reversing the change of variable yields
\begin{equation}
    \alpha = \alpha_0 e^{-i\omega_c t} - i\frac{\gamma}{\hbar} \int_0^t d\tau e^{i\omega_c(\tau-t)} \sum_i  \boldsymbol S_i(\tau) \cdot \boldsymbol B_{\rm{rms}}(\boldsymbol r_i) \,.
\end{equation}
Now, by noting that $\boldsymbol S_i^* = \boldsymbol S_i$, summing $\alpha + \alpha^*$ yields
\begin{equation}
    \alpha(t) + \alpha^*(t) = 2 \Re \left(\alpha_0 e^{-i\omega_c t}\right) + \frac{2 \gamma}{\hbar}  \int_0^t d\tau \sin(\omega_c (\tau - t)) \sum_i \boldsymbol S_i(\tau) \cdot \boldsymbol B_{\rm{rms}}(\boldsymbol r_i) \,,
    \label{eq:sola}
\end{equation}
with $\alpha_0 = \alpha(t=0)$ the initial condition of $\alpha$. Substituting Eq. \eqref{eq:sola} into Eq. \eqref{eq:eqmotionspin} yields $\dot{\boldsymbol S_i}=-\gamma\boldsymbol S_i\times\boldsymbol B_{\rm{eff}}'(\boldsymbol r_i)$, with $\boldsymbol B_{\rm{eff}}'(\boldsymbol r_i) = \boldsymbol B_{\rm{eff}}(\boldsymbol r_i) + \boldsymbol B_{\rm{cav}}(\boldsymbol r_i)$, $\boldsymbol B_{\rm{cav}}(\boldsymbol r_i) = \boldsymbol B_{\rm{rms}}(\boldsymbol r_i) \Gamma(t)$ and
\begin{equation}
    \Gamma(t) = 2 \Re \left(\alpha_0 e^{-i\omega_c t}\right) + \frac{2 \gamma}{\hbar} \int_0^t d\tau \sin(\omega_c (\tau - t)) \sum_i \boldsymbol S_i(\tau) \cdot \boldsymbol B_{\rm{rms}}(\boldsymbol r_i) \,.
    \label{eq:gammaoriginal}
\end{equation}
We introduce dissipation in the cavity by promoting $\omega_c \to \omega_c - i \kappa$ in Eq. \eqref{eq:eqmotiona}. This corresponds to local dissipation at the level of a Linblad master equation for the cavity. As a result, Eq. \eqref{eq:gammaoriginal} becomes
\begin{equation}
    \Gamma(t) = 2 e^{- \kappa t} \Re \left(\alpha_0 e^{-i\omega_c t}\right) + \frac{2 \gamma}{\hbar}  \int_0^t d \tau e^{\kappa (\tau - t)} \sin(\omega_c (\tau - t)) \sum_i  \boldsymbol S_i(\tau) \cdot \boldsymbol B_{\rm{rms}}(\boldsymbol r_i)  \,.
\end{equation}
Finally, we can express $\Gamma(t)$ in terms of the reduced magnetization, which is Mumax3's natural variable, to yield Eq. \eqref{eq:gammafinal}.

\section{Source code changes introduced in Mumax3-cQED}
\label{app:source}

The core implementation of the cavity effect in Mumax3-cQED is contained in the file \italic{cuda/cavity.go}. This file implements the cavity field described in Secs. \ref{sec:llg-plus-cavity} and \ref{sec:recursive memory term}. To interface \italic{cuda/cavity.go} with the rest of Mumax3, several other files have been created or modified:
\begin{itemize}[-]
\item \italic{cuda/cavity.go}: File to compute the cavity field, bridge between GPU and CPU (new file).
\item \italic{engine/cavity.go}: File to add the cavity field to the effective field (new file).
\item \italic{engine/effectivefield.go}: File to compute the effective field (modified file).
\item \italic{engine/run.go}: File to handle the software execution (modified file).
\item \italic{cmd/mumax3/main.go}: Starting file (modified file).
\item \italic{engine/util\_extension.go}: Utility extension file (new file).
\item \italic{cuda/Makefile}: UNIX script to compile CUDA files (modified file).
\item \italic{cuda/make.ps1}: Windows script to compile CUDA files (new file).
\item \italic{cuda/realclean.ps1}: Windows script to clean compilation files for CUDA (new file).
\end{itemize}
Besides the implementation of the cavity field, we also include a new function to log a summary of the input script along with some execution data in \italic{util\_extension.go}.
We also include some Windows script files: \italic{make.ps1} and \italic{realclean.ps1}. 

\section{Simulations of \texorpdfstring{$\boldsymbol B_{\rm{rms}}(\boldsymbol r_i)$}{B{\rinferior\minferior\sinferior}(r\iinferior)}}
\label{app:brms}

The spatial distribution of zero-point field fluctuations created by the resonator at the sample's position is calculated numerically. This is typically done by assuming a $\emph{constant}$ zero-point current $i_{\rm rms}$ that creates $\boldsymbol B_{\rm{rms}}(\boldsymbol r_i) $ at the field anti-node. 
Notice that $i_{\rm rms}$ is also assumed to be constant in time since time-dependence is given by the photon operators $\hat a$ and $\hat a^\dagger$ in the Hamiltonian of Eq. \ref{eq:startH}. 

In the case of the three-dimensional re-entrant cavity, we use COMSOL. The magnitude of $B_{\rm{rms}}$ is  chosen so to  reproduce the results by Goryachev et al.\cite{Goryachev2014}. For the bright mode ($f_{\uparrow\downarrow} = 20.8$ GHz), we set two antiparallel equal currents of $i_{\rm rms}=785$ nA and calculate the resulting spatial distribution of $\boldsymbol B_{\rm{rms}}(\boldsymbol r_i)$ (see Fig. \ref{fig:Fig3}c). We recall that the latter will be larger than the actual field produced by the cavity by a factor of $(800)^{3/2}$. This accounts for the fact that the sphere's radius in the simulation is $800$ times smaller than in the experiment.  Inserting the resulting $\boldsymbol B_{\rm{rms}}(\boldsymbol r_i)$ into mumax3-cQED (as explained below) we obtain a peak splitting at resonance equal to $2g_{\uparrow\downarrow}/2\pi \sim 2$ GHz. 
For the dark mode, we scale down the intensity of zero-point current fluctuations by a factor proportional to the resonance frequency $f_{\uparrow\uparrow} = 13.2$ GHz. This results into two parallel equal currents of $i_{\rm rms}=498$ nA that yield the spatial distribution of $\boldsymbol B_{\rm{rms}}(\boldsymbol r_i)$ shown in Fig. \ref{fig:Fig3}d. 


In the case of the superconducting cpw resonator, we use the software 3D-MLSI. This code solves the London equations to calculate the space-dependent distribution of magnetic fields created by the resulting supercurrents\cite{Khapaev2002}. 
One can estimate the zero-point current at the field antinode from the mode frequency $\omega_{\rm 0}$ and the impedance of the circuit $Z_0$ using the formula derived in Ref.\citenum{Jenkins2013}:
\begin{eqnarray}
i_{\rm rms}=\omega_{\rm 0}\sqrt{\frac{\hbar\pi}{4 Z_0}},
\label{irms}
\end{eqnarray}
We use $\omega_{\rm 0}/2 \pi = 1.4$ GHz and $Z_0 = 50$ $\Omega$ yielding $i_{\rm rms}=11.3$ nA. This produces zero-point field fluctuations that strongly depend on the line-width of the central transmission line $w$ as shown in the inset of Fig. \ref{fig:Fig5}a. 

Once the spatial distribution of $\boldsymbol B_{\rm{rms}}(\boldsymbol r_i)$ is obtained, we convert it to OVF format (\lstinline|brmsfile.ovf|) and pass it to the script.
  
\section{Micromagnetic simulations with Mumax3-cQED}
\label{app:micromag}

Simulations presented here are based on two materials widely used in cavity magnonics. YIG is a ferrimagnet famous for its record low Gilbert damping parameter $\alpha \sim 10^{-3} - 10^{-5}$\cite{Serga2010}. We set the saturation magnetization to $M_{\rm sat} = 0.14 \times 10^6$ A/m, the exchange stiffness $A = 0.37 \times 10^{-11}$ J/m and $\alpha=  10^{-4}$. On the other hand, Permalloy yields shorter spin wave lifetime but has the advantage of being easy to evaporate on different substrates. In the simulations we set $M_{\rm sat} = 0.86 \times 10^6$ A/m, $A = 1.3 \times 10^{-11}$ J/m and $\alpha=  1 \times 10^{-2}$. 


The action of the cavity is contained in the previously calculated distribution of $\boldsymbol B_{\rm{rms}}(\boldsymbol r_i)$ given in the file \lstinline|brmsfile.ovf|. On the other hand, to calculate the polariton dynamics we apply a space- and time-dependent excitation magnetic field. A drawback of the current implementation of Mumax3-cQED is that one cannot excite the system by driving the cavity, since the latter is eliminated as a dynamical degree of freedom. This prevents us from directly computing the cavity transmission. To mitigate this effect, we drive the ferromagnet with an excitation field that mimics the one created by the cavity. For this purpose, we use the same OVF file (\lstinline|brmsfile.ovf|) multiplied by a proportionality factor that accounts for the intensity of the excitation magnetic field (much larger than the zero-point field fluctuations). Finally, we multiply it by the time-dependent function ${\rm sinc}  (\omega_{\rm cutoff} t)$. This is equivalent to exciting all spin-waves at frequencies below $\omega_{\rm cutoff}$. 


As an example, we provide here the script used to obtain the data presented in Fig.  \ref{fig:Fig5}b:

\begin{lstlisting}[language=C++]
Nx := 128
Ny := 128
Nz := 32
sizeX := 800e-9 
sizeY := 800e-9
sizeZ := 150e-9 
SetGridSize(Nx, Ny, Nz)
SetCellSize(sizeX/Nx, sizeY/Ny, sizeZ/Nz)
SetGeom(Cylinder(800e-09, 150e-09))

Msat  = 0.860e6      
Aex   = 1.3e-11   
alpha = 1e-2      
Temp = 0

m = Vortex(1, 1)
B_ext = vector(0.0, 0.0, 0.0)
relax()   

X0 = 0
P0 = 0
Wc = 1.39E9 * 2 * pi
Kappa = 1e-3 * 1.39E9 * 2 * pi

B_rms = vector(0, 0, 0)
B_rms.Add(LoadFile("brmsfile.ovf"))

amp := 1e6
w_cutoff := 10e9 * 2 * pi   
simutime := 0.5e-6

B_ext.Add(LoadFile("brmsfile.ovf"), amp*sin(t*w_cutoff)/(t*w_cutoff))
run(simutime)
\end{lstlisting}

The broadband dynamic response of the ferromagnet is finally obtained by calculating the numerical Fourier transform of the resulting time-dependent spatially-averaged magnetization along the relevant direction ($\bf x$ or $\bf y$ in the previous example).

%% file: Manuscript.bbl
\begin{thebibliography}{10}
\expandafter\ifx\csname url\endcsname\relax
  \def\url#1{\texttt{#1}}\fi
\expandafter\ifx\csname urlprefix\endcsname\relax\def\urlprefix{URL }\fi
\expandafter\ifx\csname href\endcsname\relax
  \def\href#1#2{#2} \def\path#1{#1}\fi

\bibitem{Walther2006}
H.~Walther, B.~T.~H. Varcoe, B.-G. Englert, T.~Becker, Cavity quantum electrodynamics, Reports on Progress in Physics 69~(5) (2006) 1325--1382.
\newblock \href {https://doi.org/10.1088/0034-4885/69/5/r02} {\path{doi:10.1088/0034-4885/69/5/r02}}.

\bibitem{Blais2021}
A.~Blais, A.~L. Grimsmo, S.~Girvin, A.~Wallraff, Circuit quantum electrodynamics, Reviews of Modern Physics 93~(2) (2021) 025005.
\newblock \href {https://doi.org/10.1103/revmodphys.93.025005} {\path{doi:10.1103/revmodphys.93.025005}}.

\bibitem{Purcell1946}
E.~M. Purcell, H.~C. Torrey, R.~V. Pound, Resonance absorption by nuclear magnetic moments in a solid, Physical Review 69~(1–2) (1946) 37--38.
\newblock \href {https://doi.org/10.1103/physrev.69.37} {\path{doi:10.1103/physrev.69.37}}.

\bibitem{Brune1994}
M.~Brune, P.~Nussenzveig, F.~Schmidt-Kaler, F.~Bernardot, A.~Maali, J.~M. Raimond, S.~Haroche, From lamb shift to light shifts: Vacuum and subphoton cavity fields measured by atomic phase sensitive detection, Physical Review Letters 72~(21) (1994) 3339--3342.
\newblock \href {https://doi.org/10.1103/physrevlett.72.3339} {\path{doi:10.1103/physrevlett.72.3339}}.

\bibitem{Johansson2006}
J.~Johansson, S.~Saito, T.~Meno, H.~Nakano, M.~Ueda, K.~Semba, H.~Takayanagi, Vacuum rabi oscillations in a macroscopic superconducting qubit lc-oscillator system, Physical Review Letters 96~(12) (2006) 127006.
\newblock \href {https://doi.org/10.1103/physrevlett.96.127006} {\path{doi:10.1103/physrevlett.96.127006}}.

\bibitem{Hepp1973}
K.~Hepp, E.~H. Lieb, On the superradiant phase transition for molecules in a quantized radiation field: the dicke maser model, Annals of Physics 76~(2) (1973) 360--404.
\newblock \href {https://doi.org/10.1016/0003-4916(73)90039-0} {\path{doi:10.1016/0003-4916(73)90039-0}}.

\bibitem{Basov2020}
D.~N. Basov, A.~Asenjo-Garcia, P.~J. Schuck, X.~Zhu, A.~Rubio, Polariton panorama, Nanophotonics 10~(1) (2020) 549--577.
\newblock \href {https://doi.org/10.1515/nanoph-2020-0449} {\path{doi:10.1515/nanoph-2020-0449}}.

\bibitem{Jiang2023}
Z.~Jiang, J.~Lim, Y.~Li, W.~Pfaff, T.-H. Lo, J.~Qian, A.~Schleife, J.-M. Zuo, V.~Novosad, A.~Hoffmann, Integrating magnons for quantum information, Applied Physics Letters 123~(13) (Sep. 2023).
\newblock \href {https://doi.org/10.1063/5.0157520} {\path{doi:10.1063/5.0157520}}.

\bibitem{ZareRameshti2022}
B.~Zare~Rameshti, S.~Viola~Kusminskiy, J.~A. Haigh, K.~Usami, D.~Lachance-Quirion, Y.~Nakamura, C.-M. Hu, H.~X. Tang, G.~E. Bauer, Y.~M. Blanter, Cavity magnonics, Physics Reports 979 (2022) 1--61.
\newblock \href {https://doi.org/10.1016/j.physrep.2022.06.001} {\path{doi:10.1016/j.physrep.2022.06.001}}.

\bibitem{LachanceQuirion2019}
D.~Lachance-Quirion, Y.~Tabuchi, A.~Gloppe, K.~Usami, Y.~Nakamura, Hybrid quantum systems based on magnonics, Applied Physics Express 12~(7) (2019) 070101.
\newblock \href {https://doi.org/10.7567/1882-0786/ab248d} {\path{doi:10.7567/1882-0786/ab248d}}.

\bibitem{Chumak2015}
A.~V. Chumak, V.~Vasyuchka, A.~Serga, B.~Hillebrands, Magnon spintronics, Nature Physics 11~(6) (2015) 453--461.
\newblock \href {https://doi.org/10.1038/nphys3347} {\path{doi:10.1038/nphys3347}}.

\bibitem{Bondarenko2023}
A.~Bondarenko, M.~Kounalakis, S.~V. Kusminskiy, G.~Bauer, Y.~M. Blanter, Resonant magnetoelastic coupling between magnetic vortex and lattice breathing modes, in: 2023 IEEE International Magnetic Conference - Short Papers (INTERMAG Short Papers), Vol.~2, IEEE, 2023, pp. 1--2.
\newblock \href {https://doi.org/10.1109/intermagshortpapers58606.2023.10228659} {\path{doi:10.1109/intermagshortpapers58606.2023.10228659}}.

\bibitem{Lambert2019}
N.~J. Lambert, A.~Rueda, F.~Sedlmeir, H.~G.~L. Schwefel, Coherent conversion between microwave and optical photons—an overview of physical implementations, Advanced Quantum Technologies 3~(1) (Dec. 2019).
\newblock \href {https://doi.org/10.1002/qute.201900077} {\path{doi:10.1002/qute.201900077}}.

\bibitem{Hisatomi2016}
R.~Hisatomi, A.~Osada, Y.~Tabuchi, T.~Ishikawa, A.~Noguchi, R.~Yamazaki, K.~Usami, Y.~Nakamura, Bidirectional conversion between microwave and light via ferromagnetic magnons, Physical Review B 93~(17) (2016) 174427.
\newblock \href {https://doi.org/10.1103/physrevb.93.174427} {\path{doi:10.1103/physrevb.93.174427}}.

\bibitem{Barbieri1989}
R.~Barbieri, M.~Cerdonio, G.~Fiorentini, S.~Vitale, Axion to magnon conversion. a scheme for the detection of galactic axions, Physics Letters B 226~(3–4) (1989) 357--360.
\newblock \href {https://doi.org/10.1016/0370-2693(89)91209-4} {\path{doi:10.1016/0370-2693(89)91209-4}}.

\bibitem{Flower2019}
G.~Flower, J.~Bourhill, M.~Goryachev, M.~E. Tobar, Broadening frequency range of a ferromagnetic axion haloscope with strongly coupled cavity–magnon polaritons, Physics of the Dark Universe 25 (2019) 100306.
\newblock \href {https://doi.org/10.1016/j.dark.2019.100306} {\path{doi:10.1016/j.dark.2019.100306}}.

\bibitem{Li2022}
Y.~Li, V.~G. Yefremenko, M.~Lisovenko, C.~Trevillian, T.~Polakovic, T.~W. Cecil, P.~S. Barry, J.~Pearson, R.~Divan, V.~Tyberkevych, C.~L. Chang, U.~Welp, W.-K. Kwok, V.~Novosad, Coherent coupling of two remote magnonic resonators mediated by superconducting circuits, Physical Review Letters 128~(4) (2022) 047701.
\newblock \href {https://doi.org/10.1103/physrevlett.128.047701} {\path{doi:10.1103/physrevlett.128.047701}}.

\bibitem{Zhang2015}
X.~Zhang, C.-L. Zou, N.~Zhu, F.~Marquardt, L.~Jiang, H.~X. Tang, Magnon dark modes and gradient memory, Nature Communications 6~(1) (Nov. 2015).
\newblock \href {https://doi.org/10.1038/ncomms9914} {\path{doi:10.1038/ncomms9914}}.

\bibitem{GonzalezGutierrez2024}
C.~A. González-Gutiérrez, D.~García-Pons, D.~Zueco, M.~J. Martínez-Pérez, Scanning spin probe based on magnonic vortex quantum cavities, ACS Nano 18~(6) (2024) 4717--4725.
\newblock \href {https://doi.org/10.1021/acsnano.3c06704} {\path{doi:10.1021/acsnano.3c06704}}.

\bibitem{Asadchy2020}
V.~S. Asadchy, M.~S. Mirmoosa, A.~Díaz-Rubio, S.~Fan, S.~A. Tretyakov, Tutorial on electromagnetic nonreciprocity and its origins, Proceedings of the IEEE 108~(10) (2020) 1684--1727.
\newblock \href {https://doi.org/10.1109/jproc.2020.3012381} {\path{doi:10.1109/jproc.2020.3012381}}.

\bibitem{Wang2019}
Y.-P. Wang, J.~Rao, Y.~Yang, P.-C. Xu, Y.~Gui, B.~Yao, J.~You, C.-M. Hu, Nonreciprocity and unidirectional invisibility in cavity magnonics, Physical Review Letters 123~(12) (2019) 127202.
\newblock \href {https://doi.org/10.1103/physrevlett.123.127202} {\path{doi:10.1103/physrevlett.123.127202}}.

\bibitem{Cao2019}
Y.~Cao, P.~Yan, Exceptional magnetic sensitivity of $\mathcal{P}\mathcal{T}$-symmetric cavity magnon polaritons, Physical Review B 99~(21) (2019) 214415.
\newblock \href {https://doi.org/10.1103/physrevb.99.214415} {\path{doi:10.1103/physrevb.99.214415}}.

\bibitem{Zhang2017}
D.~Zhang, X.-Q. Luo, Y.-P. Wang, T.-F. Li, J.~Q. You, Observation of the exceptional point in cavity magnon-polaritons, Nature Communications 8~(1) (Nov. 2017).
\newblock \href {https://doi.org/10.1038/s41467-017-01634-w} {\path{doi:10.1038/s41467-017-01634-w}}.

\bibitem{Harder2018}
M.~Harder, Y.~Yang, B.~Yao, C.~Yu, J.~Rao, Y.~Gui, R.~Stamps, C.-M. Hu, Level attraction due to dissipative magnon-photon coupling, Physical Review Letters 121~(13) (2018) 137203.
\newblock \href {https://doi.org/10.1103/physrevlett.121.137203} {\path{doi:10.1103/physrevlett.121.137203}}.

\bibitem{Harder2021}
M.~Harder, B.~M. Yao, Y.~S. Gui, C.-M. Hu, Coherent and dissipative cavity magnonics, Journal of Applied Physics 129~(20) (May 2021).
\newblock \href {https://doi.org/10.1063/5.0046202} {\path{doi:10.1063/5.0046202}}.

\bibitem{Joseph2024}
A.~Joseph, J.~M.~P. Nair, M.~A. Smith, R.~Holland, L.~J. McLellan, I.~Boventer, T.~Wolz, D.~A. Bozhko, B.~Flebus, M.~P. Weides, R.~Macedo, The role of excitation vector fields and all-polarisation state control of cavity magnonics (2024).
\newblock \href {https://doi.org/10.48550/ARXIV.2405.14603} {\path{doi:10.48550/ARXIV.2405.14603}}.

\bibitem{MartinezLosadelRincon2023}
S.~Martínez-Losa~del Rincón, I.~Gimeno, J.~Pérez-Bailón, V.~Rollano, F.~Luis, D.~Zueco, M.~J. Martínez-Pérez, Measuring the magnon-photon coupling in shaped ferromagnets: Tuning of the resonance frequency, Physical Review Applied 19~(1) (Jan. 2023).
\newblock \href {https://doi.org/10.1103/physrevapplied.19.014002} {\path{doi:10.1103/physrevapplied.19.014002}}.

\bibitem{Lee2023}
O.~Lee, K.~Yamamoto, M.~Umeda, C.~W. Zollitsch, M.~Elyasi, T.~Kikkawa, E.~Saitoh, G.~E. Bauer, H.~Kurebayashi, Nonlinear magnon polaritons, Physical Review Letters 130~(4) (2023) 046703.
\newblock \href {https://doi.org/10.1103/physrevlett.130.046703} {\path{doi:10.1103/physrevlett.130.046703}}.

\bibitem{Macedo2021}
R.~Macêdo, R.~C. Holland, P.~G. Baity, L.~J. McLellan, K.~L. Livesey, R.~L. Stamps, M.~P. Weides, D.~A. Bozhko, Electromagnetic approach to cavity spintronics, Physical Review Applied 15~(2) (2021) 024065.
\newblock \href {https://doi.org/10.1103/physrevapplied.15.024065} {\path{doi:10.1103/physrevapplied.15.024065}}.

\bibitem{Bourhill2020}
J.~Bourhill, V.~Castel, A.~Manchec, G.~Cochet, Universal characterization of cavity–magnon polariton coupling strength verified in modifiable microwave cavity, Journal of Applied Physics 128~(7) (Aug. 2020).
\newblock \href {https://doi.org/10.1063/5.0006753} {\path{doi:10.1063/5.0006753}}.

\bibitem{Proskurin2019}
I.~Proskurin, R.~Macêdo, R.~L. Stamps, Microscopic origin of level attraction for a coupled magnon-photon system in a microwave cavity, New Journal of Physics 21~(9) (2019) 095003.
\newblock \href {https://doi.org/10.1088/1367-2630/ab3cb7} {\path{doi:10.1088/1367-2630/ab3cb7}}.

\bibitem{Flower2019a}
G.~Flower, M.~Goryachev, J.~Bourhill, M.~E. Tobar, Experimental implementations of cavity-magnon systems: from ultra strong coupling to applications in precision measurement, New Journal of Physics 21~(9) (2019) 095004.
\newblock \href {https://doi.org/10.1088/1367-2630/ab3e1c} {\path{doi:10.1088/1367-2630/ab3e1c}}.

\bibitem{MartinezPerez2018}
M.~J. Martínez-Pérez, D.~Zueco, Strong coupling of a single photon to a magnetic vortex, ACS Photonics 6~(2) (2018) 360--367.
\newblock \href {https://doi.org/10.1021/acsphotonics.8b00954} {\path{doi:10.1021/acsphotonics.8b00954}}.

\bibitem{Goryachev2014}
M.~Goryachev, W.~G. Farr, D.~L. Creedon, Y.~Fan, M.~Kostylev, M.~E. Tobar, High-cooperativity cavity qed with magnons at microwave frequencies, Physical Review Applied 2~(5) (2014) 054002.
\newblock \href {https://doi.org/10.1103/physrevapplied.2.054002} {\path{doi:10.1103/physrevapplied.2.054002}}.

\bibitem{Soykal2010}
{\"O}.~O. Soykal, M.~E. Flatté, Strong field interactions between a nanomagnet and a photonic cavity, Physical Review Letters 104~(7) (2010) 077202.
\newblock \href {https://doi.org/10.1103/physrevlett.104.077202} {\path{doi:10.1103/physrevlett.104.077202}}.

\bibitem{Yu2021}
H.~Yu, J.~Xiao, H.~Schultheiss, Magnetic texture based magnonics, Physics Reports 905 (2021) 1--59.
\newblock \href {https://doi.org/10.1016/j.physrep.2020.12.004} {\path{doi:10.1016/j.physrep.2020.12.004}}.

\bibitem{Leliaert2019}
J.~Leliaert, J.~Mulkers, Tomorrow’s micromagnetic simulations, Journal of Applied Physics 125~(18) (May 2019).
\newblock \href {https://doi.org/10.1063/1.5093730} {\path{doi:10.1063/1.5093730}}.

\bibitem{Valenzuela2024}
S.~O. Valenzuela, P.~Gambardella, K.~Garello, O.~Klein, J.~F. Sierra, J.~Sinova, Spintronic materials, Elsevier, 2024, Ch.~1, pp. 159--176.
\newblock \href {https://doi.org/10.1016/b978-0-323-90800-9.00229-8} {\path{doi:10.1016/b978-0-323-90800-9.00229-8}}.

\bibitem{Haroche2020}
S.~Haroche, M.~Brune, J.~M. Raimond, From cavity to circuit quantum electrodynamics, Nature Physics 16~(3) (2020) 243--246.
\newblock \href {https://doi.org/10.1038/s41567-020-0812-1} {\path{doi:10.1038/s41567-020-0812-1}}.

\bibitem{Tabuchi2014}
Y.~Tabuchi, S.~Ishino, T.~Ishikawa, R.~Yamazaki, K.~Usami, Y.~Nakamura, Hybridizing ferromagnetic magnons and microwave photons in the quantum limit, Physical Review Letters 113~(8) (2014) 083603.
\newblock \href {https://doi.org/10.1103/physrevlett.113.083603} {\path{doi:10.1103/physrevlett.113.083603}}.

\bibitem{Bourhill2016}
J.~Bourhill, N.~Kostylev, M.~Goryachev, D.~L. Creedon, M.~E. Tobar, Ultrahigh cooperativity interactions between magnons and resonant photons in a yig sphere, Physical Review B 93~(14) (2016) 144420.
\newblock \href {https://doi.org/10.1103/physrevb.93.144420} {\path{doi:10.1103/physrevb.93.144420}}.

\bibitem{Zhang2014}
X.~Zhang, C.-L. Zou, L.~Jiang, H.~X. Tang, Strongly coupled magnons and cavity microwave photons, Physical Review Letters 113~(15) (2014) 156401.
\newblock \href {https://doi.org/10.1103/physrevlett.113.156401} {\path{doi:10.1103/physrevlett.113.156401}}.

\bibitem{Rao2019}
J.~W. Rao, C.~H. Yu, Y.~T. Zhao, Y.~S. Gui, X.~L. Fan, D.~S. Xue, C.-M. Hu, Level attraction and level repulsion of magnon coupled with a cavity anti-resonance, New Journal of Physics 21~(6) (2019) 065001.
\newblock \href {https://doi.org/10.1088/1367-2630/ab2482} {\path{doi:10.1088/1367-2630/ab2482}}.

\bibitem{Boventer2019}
I.~Boventer, M.~Kläui, R.~Macêdo, M.~Weides, Steering between level repulsion and attraction: broad tunability of two-port driven cavity magnon-polaritons, New Journal of Physics 21~(12) (2019) 125001.
\newblock \href {https://doi.org/10.1088/1367-2630/ab5c12} {\path{doi:10.1088/1367-2630/ab5c12}}.

\bibitem{Gardin2024}
A.~Gardin, G.~Bourcin, J.~Bourhill, V.~Vlaminck, C.~Person, C.~Fumeaux, G.~C. Tettamanzi, V.~Castel, Engineering synthetic gauge fields through the coupling phases in cavity magnonics, Physical Review Applied 21~(6) (2024) 064033.
\newblock \href {https://doi.org/10.1103/physrevapplied.21.064033} {\path{doi:10.1103/physrevapplied.21.064033}}.

\bibitem{Bourhill2023}
J.~Bourhill, W.~Yu, V.~Vlaminck, G.~E.~W. Bauer, G.~Ruoso, V.~Castel, Generation of circulating cavity magnon polaritons, Physical Review Applied 19~(1) (2023) 014030.
\newblock \href {https://doi.org/10.1103/physrevapplied.19.014030} {\path{doi:10.1103/physrevapplied.19.014030}}.

\bibitem{Owens2022}
J.~C. Owens, M.~G. Panetta, B.~Saxberg, G.~Roberts, S.~Chakram, R.~Ma, A.~Vrajitoarea, J.~Simon, D.~I. Schuster, Chiral cavity quantum electrodynamics, Nature Physics 18~(9) (2022) 1048--1052.
\newblock \href {https://doi.org/10.1038/s41567-022-01671-3} {\path{doi:10.1038/s41567-022-01671-3}}.

\bibitem{Huebl2013}
H.~Huebl, C.~W. Zollitsch, J.~Lotze, F.~Hocke, M.~Greifenstein, A.~Marx, R.~Gross, S.~T.~B. Goennenwein, High cooperativity in coupled microwave resonator ferrimagnetic insulator hybrids, Physical Review Letters 111~(12) (2013) 127003.
\newblock \href {https://doi.org/10.1103/physrevlett.111.127003} {\path{doi:10.1103/physrevlett.111.127003}}.

\bibitem{Ghirri2023}
A.~Ghirri, C.~Bonizzoni, M.~Maksutoglu, A.~Mercurio, O.~Di~Stefano, S.~Savasta, M.~Affronte, Ultrastrong magnon-photon coupling achieved by magnetic films in contact with superconducting resonators, Physical Review Applied 20~(2) (2023) 024039.
\newblock \href {https://doi.org/10.1103/physrevapplied.20.024039} {\path{doi:10.1103/physrevapplied.20.024039}}.

\bibitem{Golovchanskiy2021}
I.~Golovchanskiy, N.~Abramov, V.~Stolyarov, A.~Golubov, M.~Y. Kupriyanov, V.~Ryazanov, A.~Ustinov, Approaching deep-strong on-chip photon-to-magnon coupling, Physical Review Applied 16~(3) (2021) 034029.
\newblock \href {https://doi.org/10.1103/physrevapplied.16.034029} {\path{doi:10.1103/physrevapplied.16.034029}}.

\bibitem{Wagle2024}
D.~Wagle, A.~Rai, M.~T. Kaffash, M.~B. Jungfleisch, Controlling magnon-photon coupling in a planar geometry, Journal of Physics: Materials 7~(2) (2024) 025005.
\newblock \href {https://doi.org/10.1088/2515-7639/ad2984} {\path{doi:10.1088/2515-7639/ad2984}}.

\bibitem{Bhoi2017}
B.~Bhoi, B.~Kim, J.~Kim, Y.-J. Cho, S.-K. Kim, Robust magnon-photon coupling in a planar-geometry hybrid of inverted split-ring resonator and yig film, Scientific Reports 7~(1) (Sep. 2017).
\newblock \href {https://doi.org/10.1038/s41598-017-12215-8} {\path{doi:10.1038/s41598-017-12215-8}}.

\bibitem{Hou2019}
J.~T. Hou, L.~Liu, Strong coupling between microwave photons and nanomagnet magnons, Physical Review Letters 123~(10) (2019) 107702.
\newblock \href {https://doi.org/10.1103/physrevlett.123.107702} {\path{doi:10.1103/physrevlett.123.107702}}.

\bibitem{Li2019}
Y.~Li, T.~Polakovic, Y.-L. Wang, J.~Xu, S.~Lendinez, Z.~Zhang, J.~Ding, T.~Khaire, H.~Saglam, R.~Divan, J.~Pearson, W.-K. Kwok, Z.~Xiao, V.~Novosad, A.~Hoffmann, W.~Zhang, Strong coupling between magnons and microwave photons in on-chip ferromagnet-superconductor thin-film devices, Physical Review Letters 123~(10) (2019) 107701.
\newblock \href {https://doi.org/10.1103/physrevlett.123.107701} {\path{doi:10.1103/physrevlett.123.107701}}.

\bibitem{Trif2024}
M.~Trif, Y.~Tserkovnyak, Cavity magnonics with domain walls in insulating ferromagnetic wires (2024).
\newblock \href {https://doi.org/10.48550/ARXIV.2401.03164} {\path{doi:10.48550/ARXIV.2401.03164}}.

\bibitem{Pan2024}
X.-F. Pan, P.-B. Li, X.-L. Hei, X.~Zhang, M.~Mochizuki, F.-L. Li, F.~Nori, Magnon-skyrmion hybrid quantum systems: Tailoring interactions via magnons, Physical Review Letters 132~(19) (2024) 193601.
\newblock \href {https://doi.org/10.1103/physrevlett.132.193601} {\path{doi:10.1103/physrevlett.132.193601}}.

\bibitem{Psaroudaki2023}
C.~Psaroudaki, E.~Peraticos, C.~Panagopoulos, Skyrmion qubits: Challenges for future quantum computing applications, Applied Physics Letters 123~(26) (Dec. 2023).
\newblock \href {https://doi.org/10.1063/5.0177864} {\path{doi:10.1063/5.0177864}}.

\bibitem{Sharma2022}
S.~Sharma, V.~S~V~Bittencourt, S.~Viola~Kusminskiy, Protocol for generating an arbitrary quantum state of the magnetization in cavity magnonics, Journal of Physics: Materials 5~(3) (2022) 034006.
\newblock \href {https://doi.org/10.1088/2515-7639/ac81f0} {\path{doi:10.1088/2515-7639/ac81f0}}.

\bibitem{Khan2021}
S.~Khan, O.~Lee, T.~Dion, C.~W. Zollitsch, S.~Seki, Y.~Tokura, J.~D. Breeze, H.~Kurebayashi, Coupling microwave photons to topological spin textures in cu$_2$oseo$_3$, Physical Review B 104~(10) (2021) l100402.
\newblock \href {https://doi.org/10.1103/physrevb.104.l100402} {\path{doi:10.1103/physrevb.104.l100402}}.

\bibitem{Liensberger2021}
L.~Liensberger, F.~X. Haslbeck, A.~Bauer, H.~Berger, R.~Gross, H.~Huebl, C.~Pfleiderer, M.~Weiler, Tunable cooperativity in coupled spin-cavity systems, Physical Review B 104~(10) (2021) l100415.
\newblock \href {https://doi.org/10.1103/physrevb.104.l100415} {\path{doi:10.1103/physrevb.104.l100415}}.

\bibitem{Hrabec2020}
A.~Hrabec, Z.~Luo, L.~J. Heyderman, P.~Gambardella, Synthetic chiral magnets promoted by the dzyaloshinskii–moriya interaction, Applied Physics Letters 117~(13) (Sep. 2020).
\newblock \href {https://doi.org/10.1063/5.0021184} {\path{doi:10.1063/5.0021184}}.

\bibitem{MartinezPerez2019}
M.~J. Martínez-Pérez, D.~Zueco, Quantum electrodynamics with magnetic textures, New Journal of Physics 21~(11) (2019) 115002.
\newblock \href {https://doi.org/10.1088/1367-2630/ab52d7} {\path{doi:10.1088/1367-2630/ab52d7}}.

\bibitem{Vansteenkiste2014}
A.~Vansteenkiste, J.~Leliaert, M.~Dvornik, M.~Helsen, F.~Garcia-Sanchez, B.~Van~Waeyenberge, The design and verification of mumax3, AIP Advances 4~(10) (Oct. 2014).
\newblock \href {https://doi.org/10.1063/1.4899186} {\path{doi:10.1063/1.4899186}}.

\bibitem{Khapaev2002}
M.~M. Khapaev, M.~Y. Kupriyanov, E.~Goldobin, M.~Siegel, Current distribution simulation for superconducting multi-layered structures, Superconductor Science and Technology 16~(1) (2002) 24--27.
\newblock \href {https://doi.org/10.1088/0953-2048/16/1/305} {\path{doi:10.1088/0953-2048/16/1/305}}.

\bibitem{RomanRoche2021}
J.~Rom{\'{a}}n-Roche, F.~Luis, D.~Zueco, Photon condensation and enhanced magnetism in cavity {QED}, Physical Review Letters 127~(16) (oct 2021).
\newblock \href {https://doi.org/10.1103/physrevlett.127.167201} {\path{doi:10.1103/physrevlett.127.167201}}.

\bibitem{RomanRoche2022}
J.~Román-Roche, D.~Zueco, \href{https://scipost.org/10.21468/SciPostPhysLectNotes.50}{{Effective theory for matter in non-perturbative cavity QED}}, SciPost Phys. Lect. Notes (2022) 50\href {https://doi.org/10.21468/SciPostPhysLectNotes.50} {\path{doi:10.21468/SciPostPhysLectNotes.50}}.
\newline\urlprefix\url{https://scipost.org/10.21468/SciPostPhysLectNotes.50}

\bibitem{carollo2021exactness}
F.~Carollo, I.~Lesanovsky, \href{https://link.aps.org/doi/10.1103/PhysRevLett.126.230601}{Exactness of mean-field equations for open dicke models with an application to pattern retrieval dynamics}, Phys. Rev. Lett. 126 (2021) 230601.
\newblock \href {https://doi.org/10.1103/PhysRevLett.126.230601} {\path{doi:10.1103/PhysRevLett.126.230601}}.
\newline\urlprefix\url{https://link.aps.org/doi/10.1103/PhysRevLett.126.230601}

\bibitem{romanroche2024linear}
J.~Román-Roche, Álvaro Gómez-León, F.~Luis, D.~Zueco, \href{https://arxiv.org/abs/2406.11957}{Linear response theory for cavity qed materials} (2024).
\newblock \href {http://arxiv.org/abs/2406.11957} {\path{arXiv:2406.11957}}.
\newline\urlprefix\url{https://arxiv.org/abs/2406.11957}

\bibitem{romanroche2024cavity}
J.~Román-Roche, Álvaro Gómez-León, F.~Luis, D.~Zueco, \href{https://arxiv.org/abs/2406.11971}{Cavity qed materials: Comparison and validation of two linear response theories at arbitrary light-matter coupling strengths} (2024).
\newblock \href {http://arxiv.org/abs/2406.11971} {\path{arXiv:2406.11971}}.
\newline\urlprefix\url{https://arxiv.org/abs/2406.11971}

\bibitem{Dicke1954}
R.~H. Dicke, Coherence in spontaneous radiation processes, Physical Review 93~(1) (1954) 99--110.
\newblock \href {https://doi.org/10.1103/physrev.93.99} {\path{doi:10.1103/physrev.93.99}}.

\bibitem{wang1973phase}
Y.~K. Wang, F.~T. Hioe, \href{https://link.aps.org/doi/10.1103/PhysRevA.7.831}{Phase transition in the dicke model of superradiance}, Phys. Rev. A 7 (1973) 831--836.
\newblock \href {https://doi.org/10.1103/PhysRevA.7.831} {\path{doi:10.1103/PhysRevA.7.831}}.
\newline\urlprefix\url{https://link.aps.org/doi/10.1103/PhysRevA.7.831}

\bibitem{Emary2003}
C.~Emary, T.~Brandes, Chaos and the quantum phase transition in the dicke model, Physical Review E 67~(6) (jun 2003).
\newblock \href {https://doi.org/10.1103/physreve.67.066203} {\path{doi:10.1103/physreve.67.066203}}.

\bibitem{kirton2018introduction}
P.~Kirton, M.~M. Roses, J.~Keeling, E.~G. Dalla~Torre, \href{http://dx.doi.org/10.1002/qute.201800043}{Introduction to the dicke model: From equilibrium to nonequilibrium, and vice versa}, Advanced Quantum Technologies 2~(1–2) (Oct. 2018).
\newblock \href {https://doi.org/10.1002/qute.201800043} {\path{doi:10.1002/qute.201800043}}.
\newline\urlprefix\url{http://dx.doi.org/10.1002/qute.201800043}

\bibitem{romanroche2023exact}
J.~Rom\'an-Roche, V.~Herr\'aiz-L\'opez, D.~Zueco, \href{https://link.aps.org/doi/10.1103/PhysRevB.108.165130}{Exact solution for quantum strong long-range models via a generalized hubbard-stratonovich transformation}, Phys. Rev. B 108 (2023) 165130.
\newblock \href {https://doi.org/10.1103/PhysRevB.108.165130} {\path{doi:10.1103/PhysRevB.108.165130}}.
\newline\urlprefix\url{https://link.aps.org/doi/10.1103/PhysRevB.108.165130}

\bibitem{McAllister2017}
B.~T. McAllister, Y.~Shen, G.~Flower, S.~R. Parker, M.~E. Tobar, Higher order reentrant post modes in cylindrical cavities and related comments., Journal of Applied Physics 122~(14) (Oct. 2017).
\newblock \href {https://doi.org/10.1063/1.4991751} {\path{doi:10.1063/1.4991751}}.

\bibitem{Shinjo2000}
T.~Shinjo, T.~Okuno, R.~Hassdorf, K.~Shigeto, T.~Ono, Magnetic vortex core observation in circular dots of permalloy, Science 289~(5481) (2000) 930--932.
\newblock \href {https://doi.org/10.1126/science.289.5481.930} {\path{doi:10.1126/science.289.5481.930}}.

\bibitem{Wachowiak2002}
A.~Wachowiak, J.~Wiebe, M.~Bode, O.~Pietzsch, M.~Morgenstern, R.~Wiesendanger, Direct observation of internal spin structure of magnetic vortex cores, Science 298~(5593) (2002) 577--580.
\newblock \href {https://doi.org/10.1126/science.1075302} {\path{doi:10.1126/science.1075302}}.

\bibitem{Guslienko2006}
K.~Y. Guslienko, X.~F. Han, D.~J. Keavney, R.~Divan, S.~D. Bader, Magnetic vortex core dynamics in cylindrical ferromagnetic dots, Physical Review Letters 96~(6) (2006) 067205.
\newblock \href {https://doi.org/10.1103/physrevlett.96.067205} {\path{doi:10.1103/physrevlett.96.067205}}.

\bibitem{Jenkins2013}
M.~Jenkins, T.~Hümmer, M.~J. Martínez-Pérez, J.~García-Ripoll, D.~Zueco, F.~Luis, Coupling single-molecule magnets to quantum circuits, New Journal of Physics 15~(9) (2013) 095007.
\newblock \href {https://doi.org/10.1088/1367-2630/15/9/095007} {\path{doi:10.1088/1367-2630/15/9/095007}}.

\bibitem{Serga2010}
A.~A. Serga, A.~V. Chumak, B.~Hillebrands, Yig magnonics, Journal of Physics D: Applied Physics 43~(26) (2010) 264002.
\newblock \href {https://doi.org/10.1088/0022-3727/43/26/264002} {\path{doi:10.1088/0022-3727/43/26/264002}}.

\end{thebibliography}
